\documentclass[
		twoside,openright,titlepage,numbers=noenddot,headinclude,
                footinclude=true,cleardoublepage=empty,
                BCOR=5mm,paper=a4,fontsize=11pt, 
                ngerman,american, 
                ]{scrreprt} 
                
\PassOptionsToPackage{eulerchapternumbers,listings,pdfspacing, subfig,beramono,eulermath,parts}{classicthesis}

\newcommand{\mySubtitle}{Honors Undergraduate Thesis \xspace  }
\newcommand{\myTitle}{A Novel Spin-Light Polarimeter for the  Electron Ion Collider\xspace}
\newcommand{\myDegree}{Bachelor of Science in Physics\xspace}
\newcommand{\myName}{Prajwal Mohanmurthy\xspace}

\newcommand{\myFaculty}{Put data here\xspace}
\newcommand{\myDepartment}{Department of Physics and Astronomy\xspace}
\newcommand{\myUni}{Mississippi State University\xspace}

\newcommand{\myTime}{Nov 25, 2012\xspace}
\newcommand{\myVersion}{version 1.0\xspace}

\newcounter{dummy} 
\providecommand{\mLyX}{L\kern-.1667em\lower.25em\hbox{Y}\kern-.125emX\@}

\usepackage{lipsum} 

\usepackage{color}

\PassOptionsToPackage{latin9}{inputenc} 
\usepackage{inputenc}
 
\usepackage{babel}

\PassOptionsToPackage{square,numbers}{natbib}
 \usepackage{natbib}
 
\PassOptionsToPackage{fleqn}{amsmath} 
 \usepackage{amsmath}
\usepackage{listings}
 
\PassOptionsToPackage{T1}{fontenc} 
\usepackage{fontenc}

\usepackage{xspace} 

\usepackage{mparhack} 

\usepackage{fixltx2e} 

\PassOptionsToPackage{smaller}{acronym} 
\usepackage{acronym} 

\PassOptionsToPackage{pdftex}{graphicx}
\usepackage{graphicx} 

\usepackage{tabularx} 
\usepackage{caption}
\usepackage{subfig}  

\usepackage{listings} 
\PassOptionsToPackage{pdftex,hyperfootnotes=false,pdfpagelabels}{hyperref}
\usepackage{hyperref}  
\hypersetup{
colorlinks=true, linktocpage=true, pdfstartpage=3, pdfstartview=FitV,
breaklinks=true, pdfpagemode=UseNone, pageanchor=true, pdfpagemode=UseOutlines,
plainpages=false, bookmarksnumbered, bookmarksopen=true, bookmarksopenlevel=1,
hypertexnames=true, pdfhighlight=/O, urlcolor=webbrown, linkcolor=RoyalBlue, citecolor=webgreen,
pdftitle={\myTitle},
pdfauthor={\textcopyright\ \myName, \myUni, \myFaculty},
pdfsubject={},
pdfkeywords={},
pdfcreator={pdfLaTeX},
pdfproducer={LaTeX with hyperref and classicthesis}
}   

\usepackage{ifthen} 
\newboolean{enable-backrefs} 
\setboolean{enable-backrefs}{false} 

\newcommand{\backrefnotcitedstring}{\relax} 
\newcommand{\backrefcitedsinglestring}[1]{(Cited on page~#1.)}
\newcommand{\backrefcitedmultistring}[1]{(Cited on pages~#1.)}
\ifthenelse{\boolean{enable-backrefs}} 
{
\PassOptionsToPackage{hyperpageref}{backref}
\usepackage{backref} 
\renewcommand*{\backref}[1]{}  
\renewcommand*{\backrefalt}[4]{
\ifcase #1 
\backrefnotcitedstring
\or
\backrefcitedsinglestring{#2}
\else
\backrefcitedmultistring{#2}
\fi}
}{\relax} 

\makeatletter
\@ifpackageloaded{babel}
{
\addto\extrasamerican{

}
\addto\extrasngerman{

}
}{\relax}
\makeatother

\usepackage{classicthesis} 

\usepackage{longtable}

\begin{document}

\frenchspacing 

\raggedbottom 

\selectlanguage{american} 


\pagenumbering{roman} 

\pagestyle{plain} 



\begin{titlepage}

\begin{addmargin}[-1cm]{-3cm}
\begin{center}
\large

\hfill
\vfill

\begingroup
\color{Maroon}A Novel Spin-Light Polarimeter\\ for the\\ Electron Ion Collider \\ \bigskip 
\endgroup

\spacedlowsmallcaps{\myName} 

\vfill

\includegraphics[width=12cm]{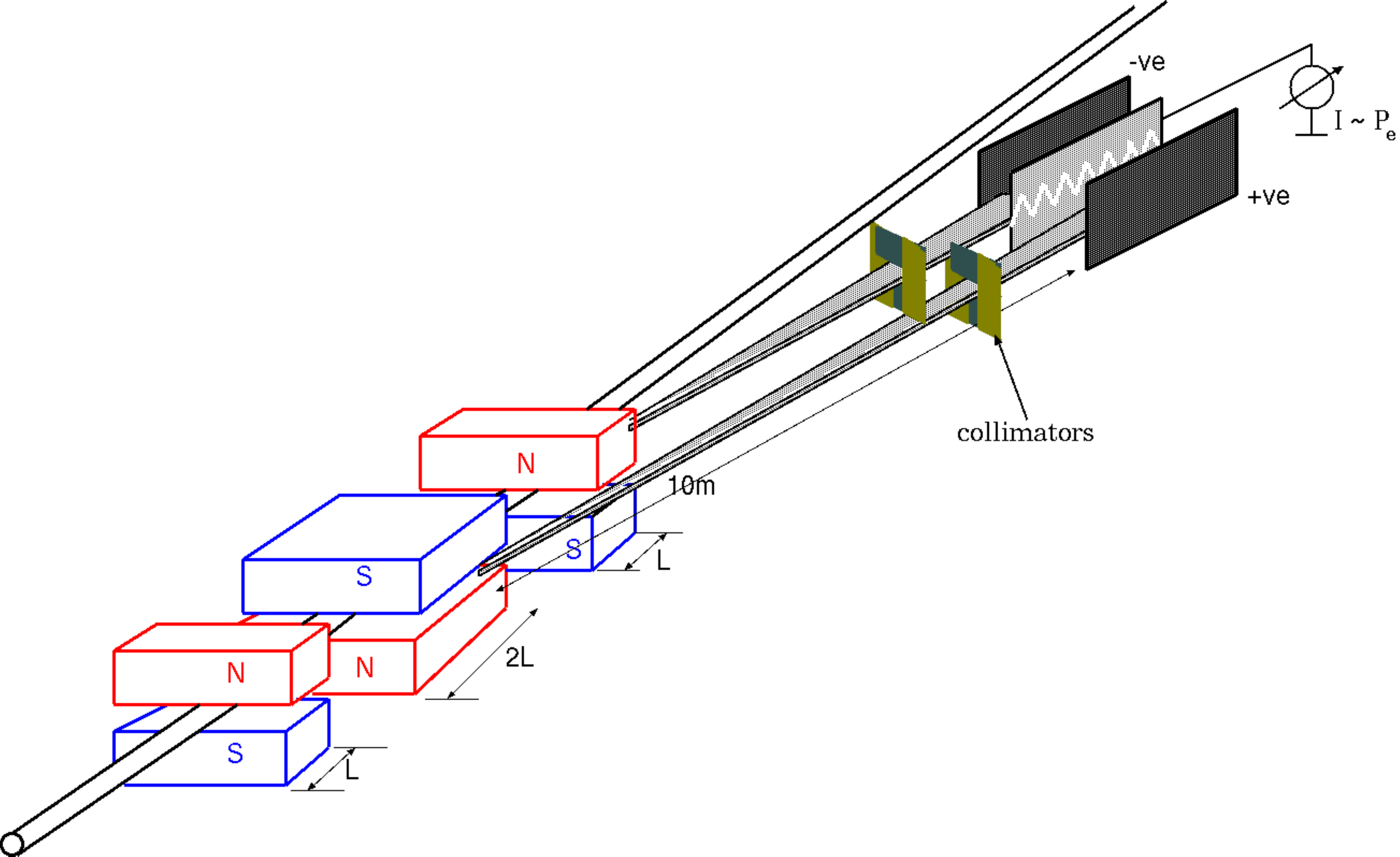} \\ \medskip 

\mySubtitle \\ \medskip 

\texttt{Submitted in partial fulfillment of the requirements for the award of the degree of}\medskip \\

\myDegree [Honors] \\
\myDepartment \\
\bigskip Mississippi State University, MS, USA

\myTime\ 

\vfill

\end{center}
\end{addmargin}

\end{titlepage} 


\thispagestyle{empty}

\newpage
\vspace{30mm}
1. Reviewer: Dr. Dipangkar Dutta

\vspace{30mm}
2. Reviewer: Dr. Seth Oppenheimer

\vspace{30mm}
3. Reviewer: Dr. Paul Reimer

\vspace{30mm}
Day of the defense: Dec 04, 2012
\hfill

\vfill

\noindent\myName:  \textit{\myTitle,} \mySubtitle, 
\textcopyright\ \myTime








\cleardoublepage

\thispagestyle{empty}
\refstepcounter{dummy}

\pdfbookmark[1]{Dedication}{Dedication} 

\vspace*{3cm}

\begin{center}
To,\\
\emph{Amma} and \emph{Appa}, \\
for their dedication and admirable way of life.

\medskip

\end{center} 

\cleardoublepage

\pdfbookmark[2]{Foreword}{Foreword} 

\begingroup
\let\clearpage\relax
\let\cleardoublepage\relax
\let\cleardoublepage\relax

\chapter*{Foreword} 
\marginpar{$^{[i]}$eRHIC: High Energy Electron-Ion collider, \url{http://www.bnl.gov/cad/eRhic/}, Retrieved on: Nov 25, 2011}
\marginpar{$^{[ii]}$ELIC: Electron Light Ion Collider at CEBAF, \url{http://casa.jlab.org/research/elic/elic.shtml}, Retrieved on: Nov 25, 2011}
$\>$ With Jefferson National Accelerator Facility's (JLAB) $12GeV$ program in construction phase with a comprehensive set of experiments already planned for the next decade, it is time to think of newer facilities that will further push the boundaries and continue the mission of a premier nuclear physics laboratory to explore the frontiers of fundamental symmetries and nature of nuclear matter. Building of an Electron Ion Collider (EIC) seems to be a natural future step. JLAB is a fixed target laboratory, but at the EIC, the target will also be accelerated thereby providing access to precision physics of quarks and gluons at much higher energies (than $12GeV$). JLAB mainly consists of the Continuous Electron Beam Accelerator and $3$ halls where the fixed target experiments are performed. Brookhaven mainly consists of the Relativistic Heavy Ion Collider with a number of main collision points on the beam line. There have been two leading proposals for the EIC, {\it i.e.}
\begin{itemize}
\item {\bf eRHIC} : Electron - Relativistic Electron Collider @ Brookhaven National Laboratory, \textit{Upton, NY} $^{[i]}$
\begin{figure}[hc]
                \centering
                \includegraphics[scale=.77]{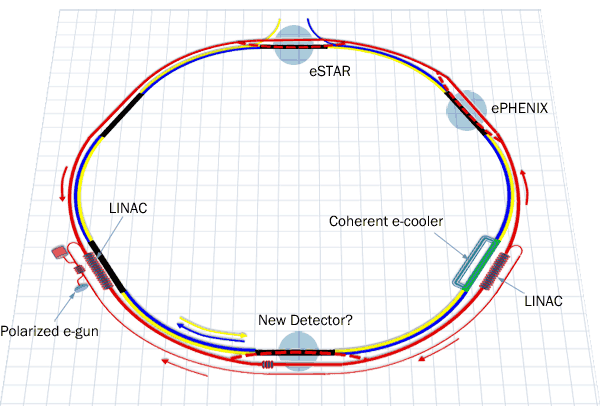}
                \label{erhic}
\end{figure}
\item {\bf ELIC} : Electron - Light Ion Collider @ Jefferson National Accelerator Laboratory, \textit{Newport News, VA} $^{[ii]}$
\begin{figure}[hc]
                \centering
                \includegraphics[scale=.5]{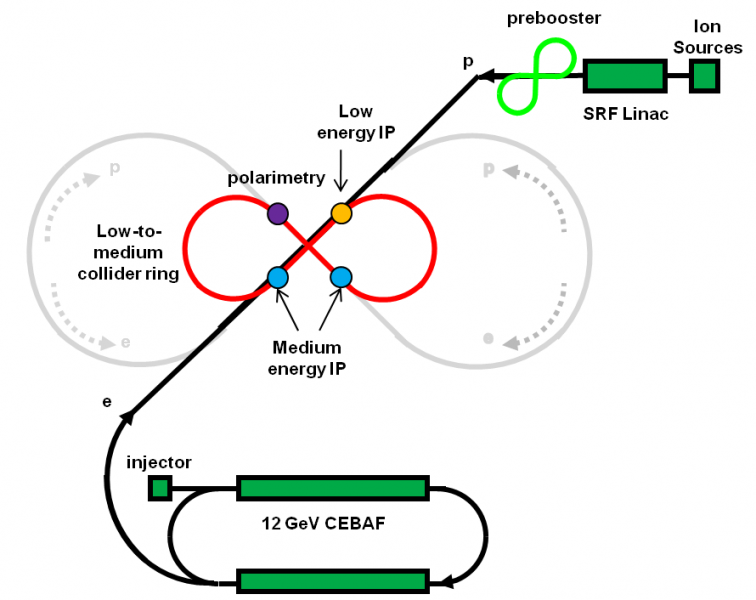}
                \label{elic}
\end{figure}
\end{itemize}
Brookhaven already has an ion accelerator and the eRHIC would need addition of an electron accelerator, whereas JLAB already has the electron accelerator and the ELIC would need addition of an ion accelerator.

At JLAB, polarization of the electron beam has played a vital role a number of experiments such as the PVDIS (Parity Violating Deep Inelastic Scattering) and the QWeak (which measured the weak charge of proton). To measure the polarization of the electron beam, JLAB has commissioned Compton and  M{\o}ller polarimeters which have met the precision demands of JLAB, but the  M{\o}ller Polarimeter generates a large background as it uses $ee$ scattering to measure the polarization. The future demands greater precision in the measurement of polarization of the beam and so at the EIC, it would be convenient to have a second non-invasive polarimeter,  besides a Compton Polarimeter, for systematics comparison.

\endgroup			

\vfill 

\cleardoublepage

\pdfbookmark[1]{Abstract}{Abstract} 

\begingroup
\let\clearpage\relax
\let\cleardoublepage\relax
\let\cleardoublepage\relax

\chapter*{Abstract} 

$\>$ A novel precision polarimeter will go a long way in satisfying
the requirements of the precision experiments being planned for a future facility such as the Electron Ion Collider. A polarimeter based on the asymmetry in the spacial distribution of the spin light component of synchrotron radiation will make for a fine addition to the existing-conventional M{\o}ller and Compton polarimeters. The spin light polarimeter consists of a set of wriggler magnet along the beam that generate synchrotron radiation. The spacial distribution of synchrotron radiation will be measured by an ionization chamber after being collimated. The up-down spacial asymmetry in the transverse plane is used to quantify the polarization of the beam. As a part of the design process, firstly, a rough calculation was drawn out to establish the validity of such an idea. Secondly, the fringe fields of the wriggler magnet was simulated using a 2-D magnetic field simulation toolkit called Poisson Superfish, which is maintained by Los Alamos National Laboratory. This was used to account for beam motion effects and the corresponding correlations were show to be negligible. Lastly, a full fledged GEANT-4 simulation was built to study the response time of the ionization chamber. This GEANT-4 simulation was analyzed for variety of effects that may hinder precision polarimetry. It was found that a Spin-Light Polarimeter would be a fine relative polarimeter.

\endgroup			

\vfill 


\cleardoublepage

\pdfbookmark[1]{Acknowledgements}{Acknowledgements} 




\begingroup

\let\clearpage\relax
\let\cleardoublepage\relax
\let\cleardoublepage\relax

\chapter*{Acknowledgements} 

This work has been generously supported by Jefferson Science Associates, LLC - Undergraduate Fellowship Program at Thomas Jefferson National Accelerator Facility, VA, USA.\\

\noindent Additional, but substantial, funding has been provided by the Mississippi State Consortium, MS, USA, and the Office of Research \& Economic Development, MS, USA. Travel funds has also been provided by the Shackouls Honors College, MS, USA, The Conference Experience for Undergraduates program of the American Physical Society's Division of Nuclear Physics, USA, and the Graduate School of Mississippi State University, MS, USA.\\

\noindent Thanks are due to the Department of Physics and Astronomy at Mississippi State University, MS, USA 
for providing office \& lab space and also the computational infrastructure required for this computational intensive work.\\

\noindent It would have been impossible to quickly adapt the GEANT4 code to the latest standards without the express support from Edward 'Jed` Legget, a graduate student in the Medium Energy Physics Group at Mississippi State University, MS, USA.\\

\noindent The author would also like to extend his gratitude to the Hall-A Compton group at Jefferson Lab, VA, USA, especially Dr. Gregg Franklin of Carnegie Mellon University, PA, USA for providing the basic Hall-A Compton chicane magnets' LANL Poisson codes.  \\

\noindent Most importantly, thanks are due to Dr. Dipangkar Dutta, who has been relentlessly at work on this project, guiding and helping the effort at every step as the major advisor and supervisor in charge.\\

\noindent Last, but not least, the author likes to thank the thesis defense committee members Dr. Paul Reimer of Argonne National Laboratory, IL, USA and Dr. Seth Oppenheimer of the Shackouls Honors College, MS, USA for their very thoughtful inputs. 

\bigskip

\endgroup 

\pagestyle{scrheadings} 

\cleardoublepage

\refstepcounter{dummy}

\pdfbookmark[1]{\contentsname}{tableofcontents} 

\setcounter{tocdepth}{2} 

\setcounter{secnumdepth}{3} 

\manualmark
\markboth{\spacedlowsmallcaps{\contentsname}}{\spacedlowsmallcaps{\contentsname}}
\tableofcontents 
\automark[section]{chapter}
\renewcommand{\chaptermark}[1]{\markboth{\spacedlowsmallcaps{#1}}{\spacedlowsmallcaps{#1}}}
\renewcommand{\sectionmark}[1]{\markright{\thesection\enspace\spacedlowsmallcaps{#1}}}

\clearpage

\begingroup 
\let\clearpage\relax
\let\cleardoublepage\relax
\let\cleardoublepage\relax

\begin{acronym}[UML]
\end{acronym}  
                   
\endgroup

\cleardoublepage 

\pagenumbering{arabic} 

\cleardoublepage 


\ctparttext{Initial work at proposal time is discussed here. Also included are some glimpses into the theory that motivates
this work. This section enumerates the work borrowed from previous work done in this field notably at Novosibirsk, Russia. 
Also presented are some recent developments in Ionization Chamber Technology.} 

\part{Ground Work} 


\chapter{Theory} 

\label{ch: Theory} 





\section{Classical SR-Power Law}

\marginpar{$^{[1]}$Proposal to the EIC R \& D:\\
\url{https://wiki.bnl.gov/conferences/images/7/74/RD2012-11\_Dutta\_eic\_polarimetry.pdf}}
\marginpar{$^{[2]}$D. D. Ivanenko, I. Pomeranchuk, Ya Zh. Eksp. Teor. Fiz. 16, 370 (1946); J. Schwinger, Phys.
Rev. 75, 1912 (1947)}
\begin{figure}[hc]
\centering
\includegraphics[scale=.4]{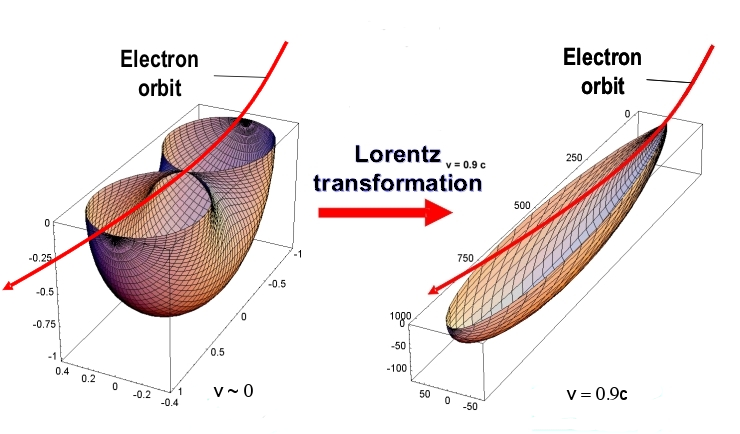}
\caption{Angular distribution of synchrotron radiation shown for the bottom half of the electron's
orbital plane. The left figure is for slow electrons, $\beta \sim 0$ and the right figure is for highly relativistic
electrons, $\beta \sim 1$. $^{[1]}$}
\label{fig1}
\end{figure}
\marginpar{$^{[3]}$I. M. Ternov and V. A. Bordovitsyn, Vestn. Mosk. Univ. Ser. Fiz. Astr. 24, 69 (1983); V. A. Bordovitsyn and V. V. Telushkin, Nucl. Inst. and Meth. B266, 3708 (2008)}

The total radiative power due to circularly accelerated particles is given by Larmor formula which is $ P_{clas} = \frac{2}{3} \frac{e^2 \gamma^4 c}{R^2}$ ({\it i.e.} $P$ is proportional to $E^4$) {\color{blue}\{here $P_{clas}$ is the classical total radiative power, $e$ is the electron charge, $\gamma = \frac{E}{m_e c^2}$ : the Lorentz boost, $c$ is the speed of light, and $R$ is the radius of trajectory of the electrons\}}. The angular dependence of radiative power can also be computed via classical electromagnetism.

\begin{equation} 
\frac{dP_{clas}}{d\Omega}  = \frac{e^2 \gamma^4 c}{4 \pi R^2}\frac{(1 - \beta Cos\theta)^2 - (1 -  \beta^2) Sin^2\theta Cos^2\phi}{(1  - \beta Cos\theta)^5}
\end{equation}

In classical electrodynamics the angular distribution of radiative power from synchrotron (SR) light can be calculated, as indicated in the above eq.(1) $^{[2]}$ {\color{blue}\{where $\mathrm{d}\Omega = \mathrm{d}\theta \mathrm{d}\phi$, and $\beta = \frac{speed of particle}{speed of light}$\}}. The SR light-cone is spread over a well defined cone with the angular spread - $\theta$ in retarded time $\Delta t'  \approx \frac{\Delta \theta}{\omega_o}$ {\color{blue}\{where $\omega_o$ is the angular frequency of the photon\}}. There is no reason to believe that the SR spectrum is mono-energetic. The spectral width of SR radiation can be formulated: $ \Delta \omega \approx \frac{1}{\Delta t' (1 - \beta)} = \frac{1}{2 \gamma^3 \omega_o}$. A critical acceleration can be envisioned at which the SR may consist of just $1$ photon by solving the equation $\gamma m_e c^2 = \hbar \omega_C$. To achieve this acceleration, a critical uniform magnetic field can be applied on an electron which is $B_{c} =  \frac{m_e^2 c^3}{e \hbar}$ {\color{blue}\{here $m_e$ refers to the rest mass of an electron\}}. The energy of the electron at these accelerations can be called the critical energy, $ E_{C} = m_e c^2 \sqrt{\frac{m_e R_c}{\hbar}} \approx 10^3 TeV$ for $R_c$ of about $10m$. It is important to note that even though accelerated electron is under consideration, the equations above don't have acceleration terms. This is because a conventional circular motion is used to calculate the parameters and only the speed is important as the acceleration is a function of speed.

\section{Quantum Corrections}

In the classical theory, the spin does not explicitly appear in the equation. But, in QED, the angular dependence of synchrotron (SR) light can be calculated to a great degree of precision and the spin of the electron involved in SR emission appears explicitly in the power law $^{[3]}$. The quantum power law for SR was worked out by Sokolov, Ternov and Klepikov as a solution to the Dirac equation $^{[4]}$ 
\marginpar{$^{[4]}$ A. A. Sokolov, N. P. Klepikov and I. M. Ternov, JETF 23, 632 (1952).}
and includes the effects introduced by electrons undergoing $ j \rightarrow j'$ (spin dependence) transitions besides also elaborating on the fluctuations to the electron orbit ($ n \rightarrow n'$ transitions - linear correction to orbit and $s \rightarrow s'$ transitions - quadratic correction to orbit). 
\marginpar{$^{[5]}$ A. A. Sokolov and I. .M. Ternov, Radiation from Relativistic Electrons, A.I.P. Translation
Series, New York (1986) ;\\ I. M. Ternov, Physics - Uspekhi 38, 409 (1995).}
The power law when integrated over all polarizations and (spacial) angular dependencies, can be written as $^{[5]}$;
\begin{equation} 
P = P_{Clas} \frac{9\sqrt3}{16\pi} \displaystyle\sum\limits_{s} \int_0^\infty \frac{y \mathrm{d}y}{(1 + \xi y)^4} I_{ss'}^2(x) F(y)
\end{equation}
\begin{multline} 
F(y) =  \frac{1 + jj'}{2} \Big[ 2 ( 1+ \xi y) \int_y^\infty K_{\frac{5}{3}}(x)\mathrm{d}x + \frac{1}{2} \xi ^2 y^2 K_{\frac{2}{3}} (y) - j(2 + \xi y) \xi y K_{\frac{1}{3}}(y) \Big] \\
+ \frac{1 - jj'}{2}\xi ^2 y^2 \Big[ K_{\frac{2}{3}}(y) + l K_{\frac{1}{3}}(y) \Big]
\end{multline}
{\color{blue}\{Where $ \xi  = \frac{3 B}{2 B_c} \gamma$, '$j$' is the spin of the electron, $y = \frac{\omega_o}{\omega_c}$, $x = \frac{3}{4}\frac{\xi \gamma^3 y^2}{(1+\xi y)^2}$, $I_{ss'}(x)$ are Laguerre functions, and $K_n(x)$ are modified Bessel functions\}}.

If $\xi << 1$ (given $B_c \approx 4.41 \times 10^9$ Tesla) then Eq. (3) can be Taylor expanded in terms of powers of $\xi$ as follows;
\begin{multline}
P  = P_{clas}\Big[\left(1 - \frac{55\sqrt3}{24}\xi + \frac{64}{3}\xi^2\right) - \left(\frac{1 + jj'}{2})(j\xi + \frac{5}{9}\xi^2 +
 \frac{245\sqrt3}{48}j \xi ^2\right) \\+  \left(\frac{1 - jj'}{2}\right)\left(\frac{4}{3}\xi^2 + \frac{315\sqrt3}{432}j\xi^2\right) + ...\Big]
\end{multline}

Eq. (3) and (4) include the a number of effects;
\begin{itemize}
\item Classical SR 
\item Thomas Precession
\item Larmor Precession
\item Interference between Larmor and Thomas Precession
\item Radiation from intrinsic magnetic moment (including anomalous magnetic moment)
\end{itemize}

Eq. (4) can be re-expressed as a difference between power from unpolarized and polarized electron beam.
\begin{equation}
P_{Spin} = P_{Pol.} - P_{UnPol.} = -j\xi P_{Clas} \int_0^\infty \frac{9\sqrt{3}}{8 \pi} y^2 K_{\frac{1}{3}}(y) \mathrm{d}y
\end{equation}
Eq. (5) is essentially the spin light that this project is based on which opens up the possibility of using SR part to measure the polarization of the beam.

The power law in eq. (5), that was derived using QED, has been extensively tested and verified at the Novosibirsk Storage ring over a large range of wavelengths (of SR). For this, the Novosibirsk group used a "snake" shaped wiggler magnet to produce the SR from a relatively low energy electron beam of about $0.5GeV @ 100\mu A$. 
\marginpar{$^{[5]}$ S. A. Belomesthnykh et al., Nucl. Inst. and Meth. 227, 173 (1984).}

\begin{figure}[hc]
\centering
\includegraphics[scale=.7]{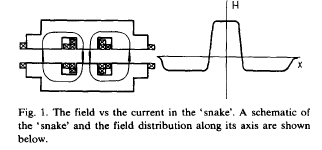}
\caption{Magnetic snake used at the VEPP-4 as a source of SR to test the spin dependence of SR. $^{[6]}$}
\label{fig2}

\end{figure}
\begin{figure}[h]
\centering
\includegraphics[scale=.6]{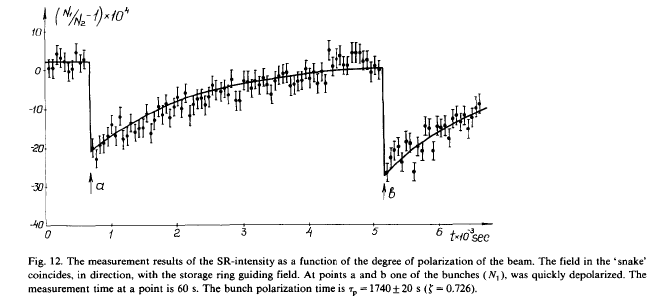}
\caption{Results from the experiment showing the increase in intensity of SR as the polarization builds up and then suddenly drops to zero when an RF field is used to depolarize the beam. $^{[6]}$}
\label{fig3}
\end{figure}
\marginpar{$^{[6]}$ K. Sato, J. of Synchrotron Rad., 8, 378 (2001).}
Figure 3, clearly demonstrated the power dependence of the SR on beam-electron polarization. It remains to be shows as to how the SR spectra could be measured.

The spin flip probability has a special significance in modern day electron storage rings and is given by the relation $^{[4]}$;
\begin{equation}
W_{\uparrow \downarrow} = \frac{1}{\tau}(1 + j \frac{8 \sqrt3}{15})
\end{equation}
\marginpar{$^{[7]}$ J. Le Duff, P. C. Marin, J. L. Manson, and M. Sommev, Orsay - Rapport Technique, 4-73
(1973).}{\color{blue}\{where $j = +1$ is spin along the magnetic field (and $j = -1$ is spin against magnetic field), and $\tau$ is the time involved in the process\}}.

As a result of the probability of spin aligning with the magnetic field direction would become very high over long periods of time. A circulating electron beam, such as ones in storage rings self polarize, and this has been studied 
in great detail at many storage rings such as ones at DESY, PSI and CESR though it was first observed at the Orsay storage ring $^{[7]}$.

\section{Spin - Light}

$\>$ In the Spin-Light polarimeter, the spin-flip term in the power law does not play an important role. The integral power law without the spin-flip term can be written as $^{[4]}$;
\begin{multline}
P_{\gamma} = \frac{9 \eta_e}{16\pi^3} \frac{c e ^2}{R ^2} \gamma^5 \int_0^{\inf} \frac{y^2 \mathrm{d}y}{ (1+\xi y)^4} \oint \mathrm{d}\Omega (1 + \alpha ^ 2)^2 \times \\
\Big[ K^2_{\frac{2}{3}}(z) + \frac{\alpha^2}{1 + \alpha^2} K^2_{\frac{1}{3}}(z) + j \xi y \frac{\alpha}{\sqrt{1 + \alpha^2}}K_{\frac{1}{3}}(z)K_{\frac{2}{3}}(z) \Big]
\end{multline}
where $ z = \frac{\omega}{2\omega_C} ( 1+\alpha^2)^{\frac{2}{3}}$ and $ \alpha = \gamma \psi$ where $\psi$ is the vertical angle {\it i.e.} above and below the orbit of the electron. Notice that the last term with a $j$ disappears from the integral over all angles ($-\frac{\pi}{2} \leq \psi \leq \frac{\pi}{2}$). But for an electron that is polarized, the power below ({\it i.e.} $-\frac{\pi}{2} \leq \psi \leq 0$ and above (i.e. $-\frac{\pi}{2} \leq \psi \leq \frac{\pi}{2}$) are spin dependent. More importantly the difference between the power radiated above and power radiated below is directly spin dependent, which can be directly obtained from Eq. (7) in differential form for circular arcs in the circular cross-section  the SR-light cone at an angle $\theta$.
\begin{multline}
\frac{\Delta P_{\gamma}(j)}{\Delta \theta} = \frac{3}{2}\frac{\hbar c \gamma^3 y}{R} \times \\
\frac{3}{\pi^2}\frac{1}{137}\frac{I_e}{e} j \xi \gamma \int_{y_1}^{y_2} y^2 \mathrm{d}y \int_0^\alpha \alpha(1 + \alpha^2)^{\frac{3}{2}} K_{\frac{1}{3}}(z)K_{\frac{2}{3}}(z) \mathrm{d}\alpha
\end{multline} 

\chapter{Initial Design} 

\label{ch: Initial Design} 

$\>$ It is obvious now that our setup will have a wiggler magnet which shall be the source of SR and an ionization chamber to measure the power spectra of the SR emitted at the wiggler magnet by the polarized electrons.

\section{Wiggler Magnet}
 
$\>$ In order to create a fan of SR light (as illustrated in Figure 5), the electron beam could be made to bend in presence of a magnet. An arrangement that would lead to the production of the SR-Cone must look similar to the "snake" magnets that
were used by VEPP-4 as described in \emph{Chapter 1}. 

\begin{figure}[h]
\centering
\includegraphics[scale=.5]{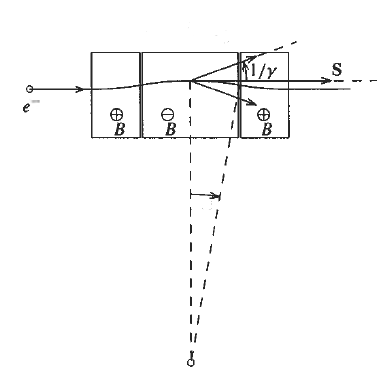}
\caption{A 3-pole wiggler with central dipole twice the length of end dipoles.}
\label{fig4}
\end{figure}

A set of 3 dipoles (3 - pole dipole), each with constant uniform magnetic field would be ideal for this purpose. The central dipole would have twice the pole length as compared to the ones on either side of the central dipole, but with a magnetic field an opposite polarity like in Figure 4.  This design will give rise to 4 fans, 2 towards either side of the beam (left and right). Each of the fan will have spacial asymmetry as a function of spin - polarization of the electrons in the vertical plane (up - down the electron beam's orbit which is perpendicular to the plane which contains the 4 fans). The 4 fans help characterize the systematics better since this configuration will flip the sign of the spin - dependent term in Eq. (8) twice essentially returning the sign to the original status.

With this geometry in mind, one could then simulate and calculate the requisite pole strengths and pole lengths appropriate to energies at Electron Ion Collider (EIC). Of course, one would also have to consider time scales at which the statistics would be sufficient to achieve the design requirement ($<1\%$) of precision. A number of techniques were employed to tackle the above issues. Including issues such as optimizing the distance between each pole were solved through a full fledged GEANT-4 simulation.

\section{Collimator}

\begin{figure}[h]
\centering
\includegraphics[scale=.3]{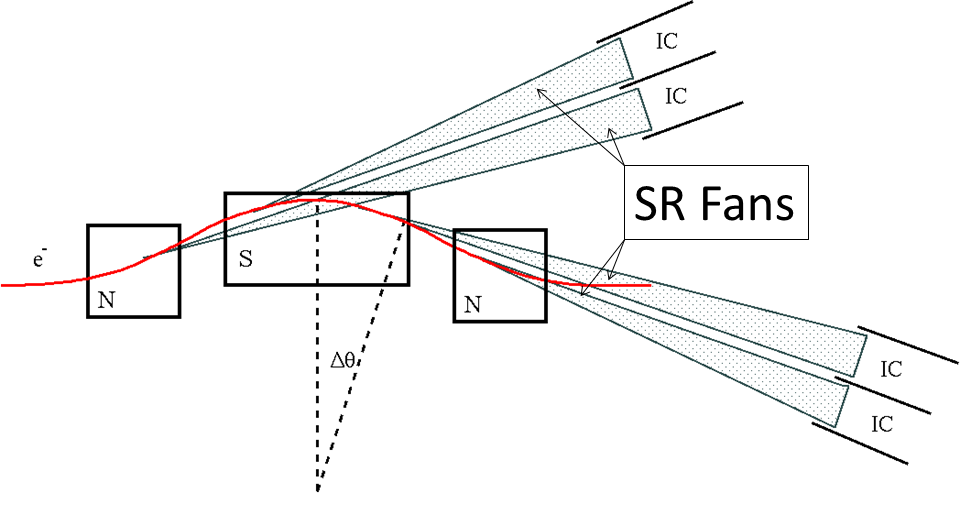}
\caption{A rough schematic of the $4$ fans that shall be created by the Wiggler Magnet}
\label{fig5}
\end{figure}

 The $4$ fans of SR-Light need not be separated from each other. They may in fact overlap and extracting the power - asymmetry information from complicated overlappings would require extensive modeling. This might also introduce new sources of uncertainties. Therefore to uniquely sample each fan at the ionization chamber, collimators would need to be placed on each face of the dipoles to direct and separate the fans of SR. The position of the collimators will be calculated using optimized values of pole - strengths, pole - lengths and relative position of each of the dipoles. 

\section{Ionization Chambers}

The Ionization Chambers (IC), one on each side ($2$ fans per IC) of the beam, could be used for measuring the spacial asymmetry in the SR - fans to be used to compute the polarization of the electrons in the beam. One would not expect a large asymmetry in the Spin-Light component (of the order of about $10^{-4}$), therefore a high resolution, low noise IC is demanded. The IC will be very close to the beam line and the spin-independent background may be as high as $ 10^{12}photons/s$, therefore the ICs have to be radiation hard. The geometry of the magnets could be changed to deal with SR Spectra with characteristic energy peak in the ranges of about $~500keV - 2.5MeV$. 
\marginpar{$^{[8]}$ A. E. Bolotnikov and B. Ramsey, Nucl. Inst. and Meth. A396, 360 (1997).}
An IC with Xenon as ionization media operated in the current mode can in principle handle high fluxes and have low noise disturbances $^{[8]}$.  It might be important to note that the Spin-Light asymmetry is spread over the entire spectra of the SR. 
\marginpar{$^{[9]}$ T. Doke, Portugal Phys. 12, 9 (1981).}
\marginpar{$^{[10]}$ V. V. Dmitrenko et al., Sov. Phys.-tech. Phys. 28, 1440 (1983); A. E. Bolotnikov et al., Sov.
Phys.-Tech. Phys. 33, 449 (1988)}
\marginpar{$^{[11]}$ C. Levin et al., Nucl. Inst. and Meth. A332, 206 (1993).}
\marginpar{$^{[12]}$ G. Tepper and J. Losee, Nucl. Inst. and Meth. A356, 339 (1995).}
\marginpar{$^{[13]}$ A. E. Bolotnikov and B. Ramsey, Nucl. Inst. and Meth. A383, 619 (1996).}
\marginpar{$^{[14]}$ G. Tepper and J. Losee, Nucl. Inst. and Meth. A368, 862 (1996).}
\marginpar{$^{[15]}$ Proportional Technologies Inc., \href{www.proportionaltech.com}{www.proportionaltech.com}.}
Sampling radiation over large energy spectrum becomes important. Since Xenon has the lowest ionization energy of about $21.9eV$, among non-radioactive nobel gases, it seems to be an ideal candidate. ICs with Xenon under high pressures have already been developed and well tested to perform well in the energy ranges of $50keV - 2.0MeV$ $^{[9]}$. Pressures involved in HPXe ICs exceed $50 atm @ 0.55g/cc$ but they work well at room - temperature $^{[10], [11],[12]}$. One of the bottlenecks was the precision of purity of the Xenon gas in its pristine form. But owing to advances in gas purification techniques $^{[13]}$, a best energy resolution of $2.4\%$ at about $0.662MeV$ has been shown to be possible when current signals from the shower is used in presence of prompt Xenon scintillation $^{[14]}$. The results from this attempt has been fairly promising (Figure 6). ICs with $3\% - 4\%$ energy resolution are even being sold commercially by \textsc{Proportional Technologies Inc.} $^{[15]}$.

\begin{figure}[h]
\centering
\includegraphics[scale=.8]{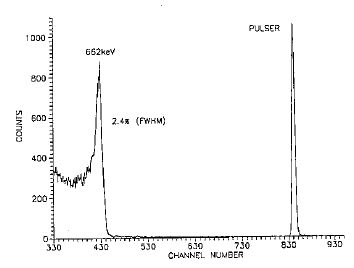}
\caption{$Cs^{137}$: $E_i = 1.7kV/cm$, pulse height spectrum in $57 atm$ at $295K$ of xenon, $E_i = 1.7kV/cm$ $^{[14]}$.}
\label{fig5}
\end{figure} 

\cleardoublepage 


\ctparttext{In this section, the work is presented with additional studies resulting from discussion with experts. It includes establishing the idea with a "back of the envelop calculation"
 backed by a full fledged GEANT-4 simulation. Also, related effects such as beam motion were studied and their effects that impact the polarimeter negatively were shown to be minimal.
Even though the GEANT-4 simulation is still in the making, a skeletal code is briefly explained here.} 

\part{Project Work} 


\chapter{Design Considerations} 

\label{ch:DesignCharacteristics} 

\section{Spin - Light Characteristics}

\begin{figure}[h]
\centering
\includegraphics[scale=.6]{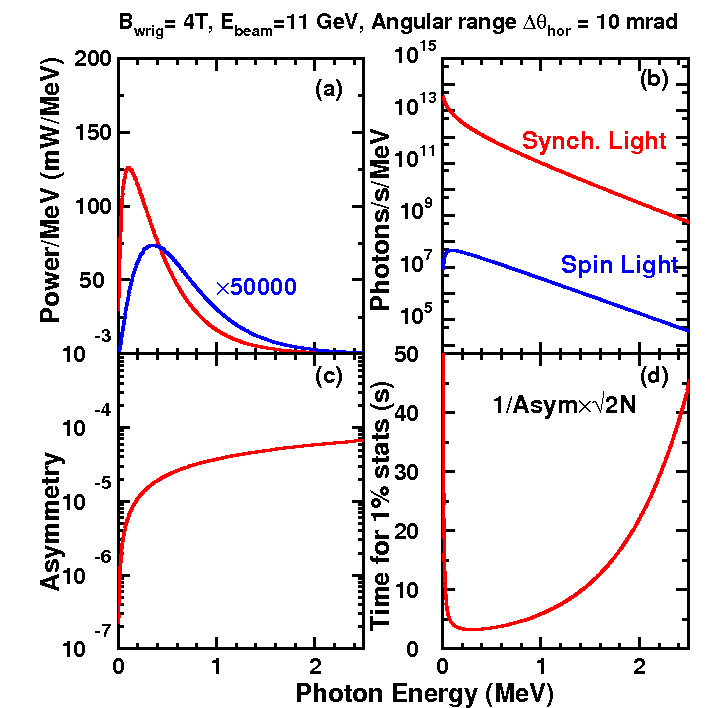}
\caption{
A. Plot of difference between the number of SR and spin light photons that go above and below the orbit of the electron ($\Delta P_{\gamma}$) vs. their energy\\ 
B. Plot of total number of SR, spin light photons $P_{\gamma}$ vs. their energy\\ 
C. Plot of the Asymmetry vs. the photon energies\\
D. Plot of time required to achieve $1\%$ statistics by sampling one wavelength of the spin-light spectra vs. the energy of the photons.}
\label{fig7}
\end{figure}

The spin - dependence of the SR can be studied by examining Eq. (6-8) in \emph{Chapter 1}. Using $I_e = 100\mu A$ and  $E_e = 11GeV$, Eq. (6-8) were numerically integrated (Appendix B.1) between $ -1< \alpha <1$ and $\Delta \theta = 10 mrad$ , for a uniform magnetic field of $B = 4T$ assuming $100\%$ longitudinal polarization.

In Figure 7, the total number of SR and spin light ($P_{\gamma}$) photons radiated is plotted. Also in Figure 7,  is a plot the difference between the power of spin light spectra above and below the orbit of the electron ($\Delta P_{\gamma}$). An asymmetry term is defined to be $A = \frac{\Delta P_{\gamma}}{P_{\gamma}}$, which was used to nail down the range of energies of the photons which must be measured. Lastly, a plot of sampling time required ($T_s = \frac{\Delta A}{A} = \frac{1}{A \sqrt{2P E_e}} $) to achieve the design precision goals.

It immediately becomes clear that the ionization chamber, which is envisioned to measure the asymmetry (that in turn will be used to compute the electron polarization), will have to be operational at wavelengths corresponding to hard - XRays. Furthermore, the asymmetry plot, Figure 7.D, demands that the sampling be done at the higher energies (and not close to ~$0.5MeV$) since at higher energies the asymmetry is not rapidly changing, thus making it an ideal high-energy polarimetry technique. It might be important to note that the asymmetry is fairly low but since the integrated power of spin-light is very high, the time required for achieving $1\%$ polarimetry is only of the order of a few seconds. Also, one could plot the asymmetry and SR spectra for different energies to study the trends with change in beam energy. This indicates that there are no suppression effects at higher energies that might hinder effective polarimetry.

\begin{figure}[h]
\centering
\includegraphics[scale=.5]{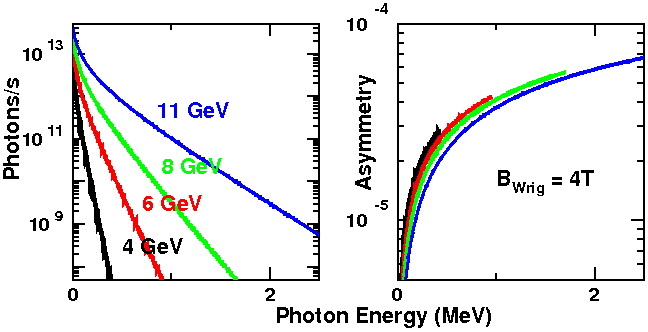}
\caption{(Left) A. Plot of spin light spectra over photon energies for various electron beam energies ranging from $4GeV - 12 GeV$; \\
(Right) B. Plot of asymmetry over photon energies for various electron beam energies ranging from $4GeV - 12 GeV$}
\label{fig8}
\end{figure}

\section{Wiggler Magnet}

\begin{figure}[h]
\centering
\includegraphics[scale=.7]{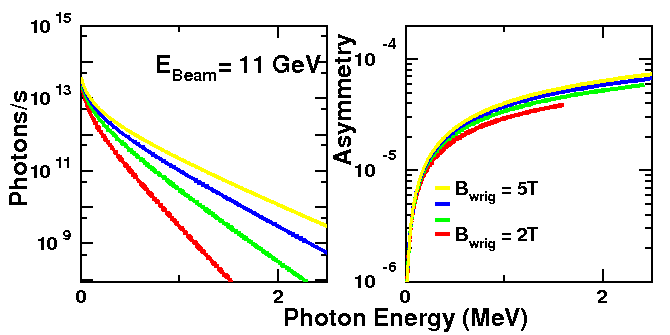}
\caption{(Left - Right)\\
A. Plot of total number of spin light photons $P_{\gamma}$ vs. their energy for various pole strengths\\ 
B. Plot of the newly defined term - Asymmetry vs. the photon energies for various pole strengths}
\label{fig9}
\end{figure}

In order to establish the dimensions of the dipoles,  the same graphs as in \emph{Section 3.1} Figure 8 were plotted for various pole lengths and magnetic fields. First, the asymmetry increases very slowly with field strength as shown in Figure 9, and the figure of merit (time for $1\%$ statistics) improves very slowly with magnetic fields above $3T$ as shown in Figure 10B, therefore $B = 4T$ was chosen since $4T$ wiggler magnets are easily available at light sources around the world. Secondly, an appropriate pole length of $L_p = 10cm$ was selected by looking at Figure 10 and selecting out the pole length for the pole strength of $4Tesla$. It is noteworthy to see that the plot of pole length as a function of pole strength was done keeping in mind an SR fan -  angular spread of about $10mrad$. The plot in Figure 10.B also re-assures the reasonable time requirement to achieve the design precision goal of $1\%$. The last parameter in the wiggler to be fixed is the distance between each dipole. 

\begin{figure}[h]
\centering
\includegraphics[scale=.7]{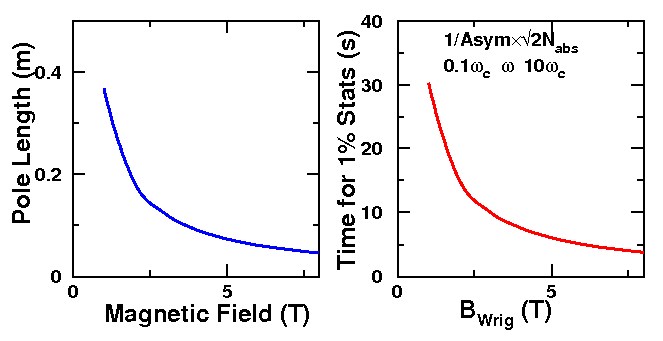}
\caption{(Left - Right)\\
A. Pole length required for a $10mrad$ angular spread of SR light fans with $B = 4T$\\ 
B. Dependence of time required to achieve $1\%$ statistics by sampling one wavelength of the spin-light spectra with pole strengths.}
\label{fig10}
\end{figure}

\subsection{Effects of wiggler on the beam}
\marginpar{$^{[16]}$ B. Norum, CEBAF Technical note, TN-0019 (1985).}
\marginpar{$^{[17]}$ M. Sands, SLAC Technical note, SLAC-121 (1970).}
A polarimeter must be non - invasive and therefore answering the question of what would be the effects of putting such as polarimeter on a beam line is very important. But the effect of a high energy electron beam emitting SR has been well studied $^{[16]}$. The number of photons ($N$) emitted by an electron when it is deviated by a radian, from its initial linear trajectory, when acted upon by a magnetic field is distributed as per the conventional Poisson distribution $^{[17]}$ about a mean value of $n$;
\begin{equation}
\bar{N}(n) = \frac{n^N e^{-n}}{N!} \\
\end{equation}
\begin{equation}
n(E_e) = \frac{5}{2\sqrt3}\frac{\gamma}{137} = 20.6E_e
\end{equation}
The average energy of the SR photons can also be written down as {\color{blue}\{where $E_e$ is the electron energy\}};
\begin{equation}
\bar{E}_e = \hbar\bar{\omega} = \frac{3}{2}\frac{\hbar c \gamma^3}{R} = \frac{3}{2}\frac{\hbar E_e^3}{R m_e^3c^5}
\end{equation}
In the case of a spin-light polarimeter, the beam energy is about $11GeV$ and we choose pole strength to be about $4T$ in \emph{Section 3.2}. An angular bend of about $10mrad$ of the beam is sufficient for such a polarimeter. Using the values of average number of photons emitted and their average energy, the average energy fluctuation ($\Delta \bar{E}_e$) of the beam can be computed.
\begin{equation}
n  = 20.62 \times 11 _{GeV} \times .01 _{rad} = 2.06
\end{equation}
\begin{equation}
\bar{E}_e =\frac{3}{2}\frac{\hbar (11_{GeV})^3}{10_m m_e c^5} = .199MeV
\end{equation}
\begin{equation}
\frac{\Delta \bar{E}_e}{E_e}  = \frac{\sqrt n \bar{E}_e}{E_e} \approx 2.5 \times 10^{-5} 
\end{equation}
The energy fluctuations are smaller than the typical precision with which the energy can be measured at an electron accelerator.

Another parameter which needs to be checked before proceeding, is the transverse kicks ($\Delta \theta_e$) received by the electrons when emitting SR photons in the magnets. The transverse kicks can be calculated in terms of angles knowing that the SR power spectrum usually peaks at an angle $\theta_\gamma = \frac{1}{\gamma}$ $^{[17]}$ {\color{blue}\{where $E_{\gamma}$ is the SR - photon energy\}};
\begin{equation}
\Delta\theta_e = \frac{E_\gamma Sin(\theta_\gamma)}{E_e} \approx 11.3 \times 10^{-9} \frac{E_{e_{(GeV)}}}{R_(m)}
\end{equation}
\begin{equation}
\bar{\theta}_e = \sqrt n \Delta\theta_e \approx 1.5 \times 10^{-8}_{(rad)}
\end{equation}
It can be clearly seen from Eq. (14) and Eq. (16) that both energy fluctuation and angular kicks shall be negligible. This can be seen for all practical purposes in the GEANT-4 simulation that this work demands. This polarimetry method remains a non-invasive procedure.

\subsection{Effects of realistic dipole magnetic field with fringes}
\begin{figure}[h]
\centering
\includegraphics[scale=.25]{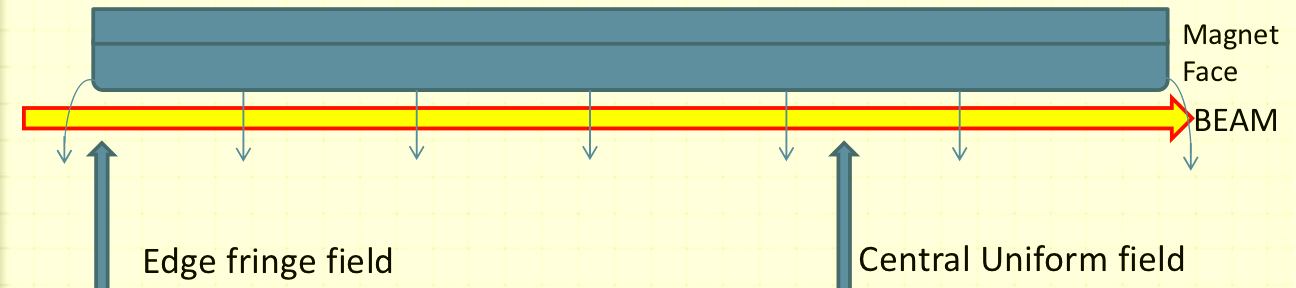}
\caption{Schematic diagram of the planes at which position the simulation was carried out.}
\label{fig11}
\end{figure}

\begin{figure}[h!]
\centering
\includegraphics[scale=.5]{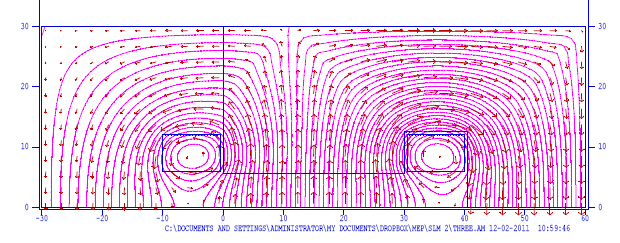}
\caption{Field map of the dipole face at the center of the dipole.}
\label{fig12}
\end{figure}
\begin{figure}[h!]
\centering
\includegraphics[scale=.3]{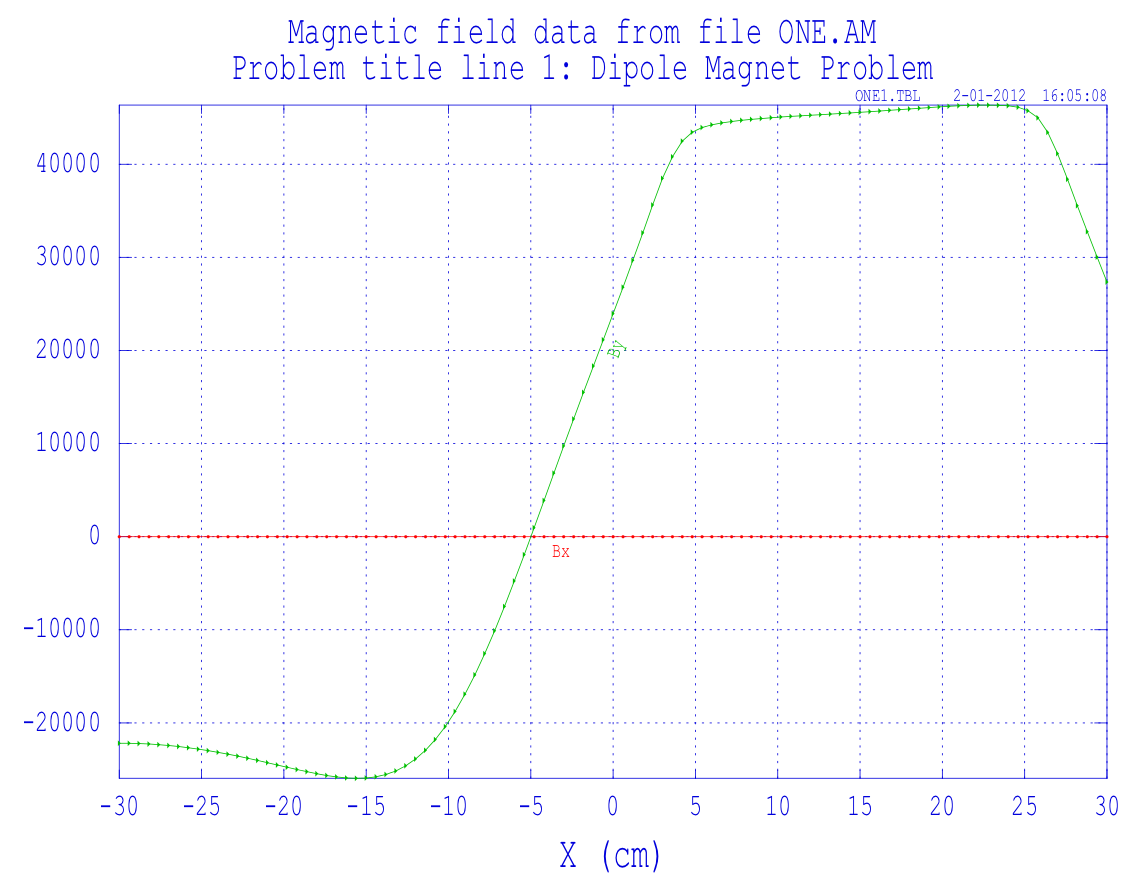}
\caption{Plot of both the $x$ and $y$ components of the magnetic field on the transverse plane at the the center of the dipole (Beam pipe is centered around $15$cm mark along the 'x' axis).}
\label{fig13}
\end{figure}

\begin{figure}[h!]
\centering
\includegraphics[scale=.5]{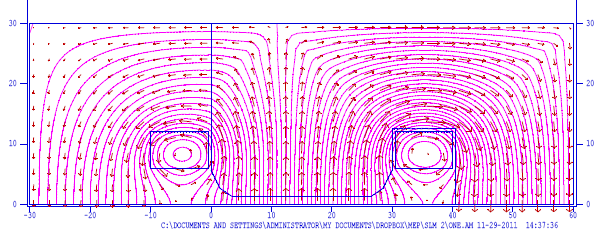}
\caption{Field map of the dipole face at the edge of the dipole.}
\label{fig14}
\end{figure}

\begin{figure}[h!]
\centering
\includegraphics[scale=.32]{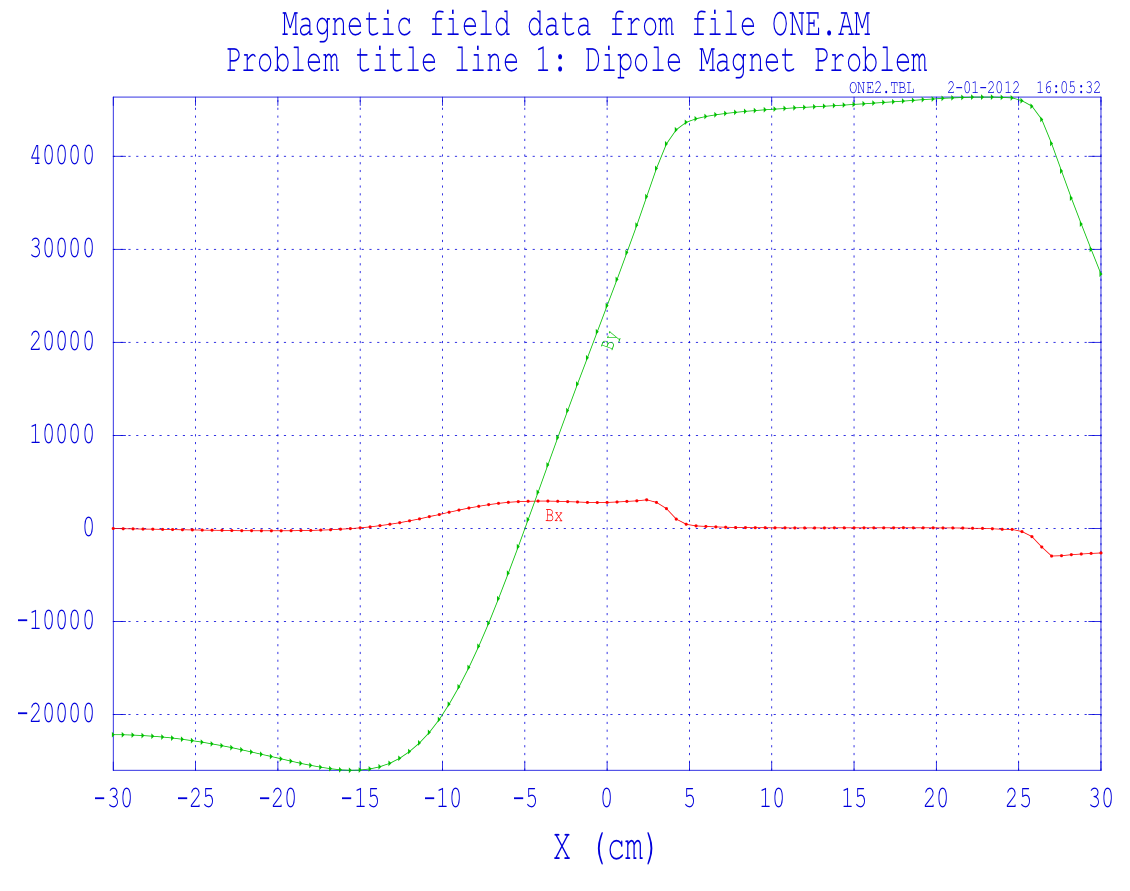}
\caption{Plot of both the $x$ and $y$ components of the magnetic field on the transverse plane at the the edge of the dipole (Beam pipe is centered around $15$cm mark along the 'x' axis).}
\label{fig15}
\end{figure}
\marginpar{$^{[18]}$ Poisson SuperFish 2D EM Solver, \url{laacg1.lanl.gov/laacg/services/sfu\_04\_04\_03.phtml},2007.}
In \emph{Section 3.2}, while plotting the power spectra and the asymmetry generated by the code in \emph{Appendix B.1} a uniform field was used. But the code can also take a field map. A field map can be generated by solving Maxwell's equations with appropriate boundary conditions. This is essential since field in the transverse plane (perpendicular to the motion of electrons) might distort the SR spectrum and thereby change the asymmetry. In fact there is a custom built suite of programs written by \textit{Los Alamos National Laboratory} to precisely do this called \textbf{LANL Poisson SuperFish} $^{[18]}$.

In LANL SuperFish, the magnet geometry can be easily defined as is done in \emph{Appendix B.2}. The field map of the magnet can then be plotted. Here, the field map at the edge where the electron beam enters the magnet and at the center of the dipole is presented. In Figures 11 \& 13, note that the beam pipe is going at the center below the magnet pole. In Figure 13, the physical taper of the cores can be notices, since it is at the edge of the magnet face. This taper of the poles is absent in Figure 11, since it is at the center. In Figures 11 \& 13, the singularities seen are the areas where the current cuts the plane. Also, it is important to note that the entire 'C' magnet is not visible in the field-map,  only the top half of the C magnet is shown in the field map.

\begin{figure}[h!]
\centering
\includegraphics[scale=.75]{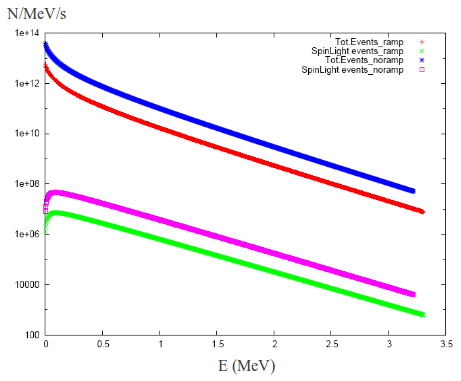}
\caption{Plot showing the SR - Light (\emph{TotEvents\_ramp}) and Spin - Light (\emph{SpinLightEvents\_ramp}) power spectra with a realistic taper for the dipoles (Power spectra for uniform magnetic field have also been presented as \emph{\_noramp}).}
\label{fig16}
\end{figure}
\begin{figure}[h!]
\centering
\includegraphics[scale=.3]{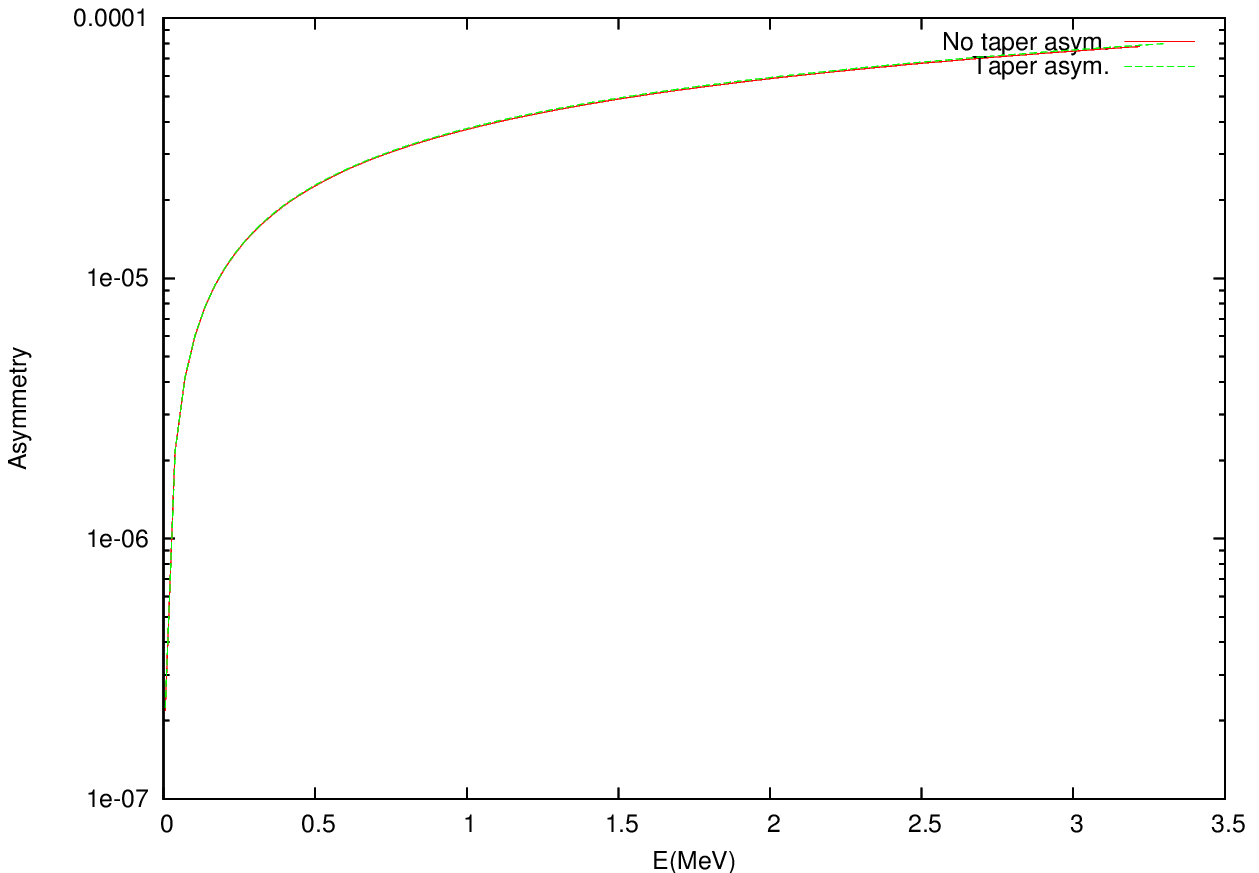}
\caption{Plot of the assymetry with a realistic taper for the dipoles (\emph{Taper asym.}).}
\label{fig17}
\end{figure}

X - Axis is to the right and left of the beam and Y - Axis is to the above and below the beam. Also the XY plane is perpendicular to the direction of motion of electron. In Figure 13, it might be important to note that there is no   component of the magnetic field. This is because it is at the center of the dipole and there is no fringing of the field. But in Figure 15, there is a non -  zero X component to the magnetic field since it is on the plane at the face of the magnet. A 2D simulation is sufficient since any components along the motion of electrons (Z Axis) will not affect the electrons.

The field map obtained here can be inserted into the numerical integration code (in \emph{Appendix B.1}) and the power spectra and the asymmetry can be obtained. Even though there is a reduction in the total power output of light by introducing a realistic taper for dipole fields, the asymmetry has not changed. This implies that the changes introduces by the realistic dipoles are minimal.

\section{Collimation and Spin-Light fan size}
$\>$ Even though the distance between the 3 dipoles should in theory not affect the physics involved, it is nevertheless an essential design parameter. A reasonable value of about $1m$ distance between each dipole was used to start with but this value will be definitely fixed with a full fledged GEANT-4 simulation. A fan with $10mrad$ spread would then give rise to a spot which is $10cm$ big in the horizontal plane, $10m$  from the wiggler where the ionization chambers will be placed. A more important dimension of the SR-spot at the ionization chamber is its height in the vertical direction. An angular spread of $\Delta \theta = 1/\gamma = 100\mu rad$ would then give rise to a spot which is $1mm$ big in the horizontal plane, $10m$  from the wiggler, where the ionization chambers will be placed.

\begin{figure}[h!]
\centering
\includegraphics[scale=.25]{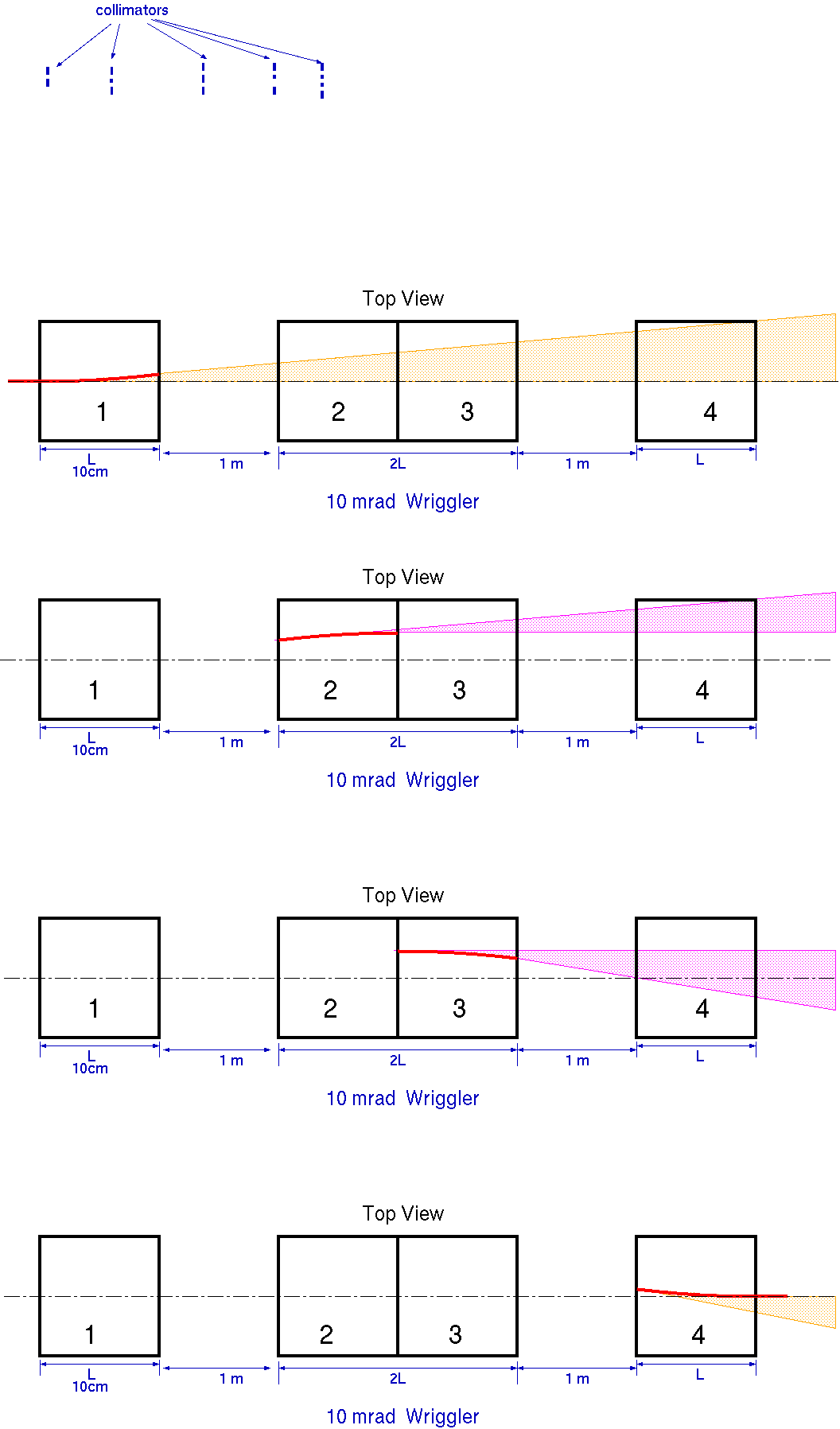}
\caption{A schematic diagram showing the 4 fans of SR that originate at the wiggler magnet system.}
\label{fig18}
\end{figure}

\begin{figure}[h!]
\centering
\includegraphics[scale=.3]{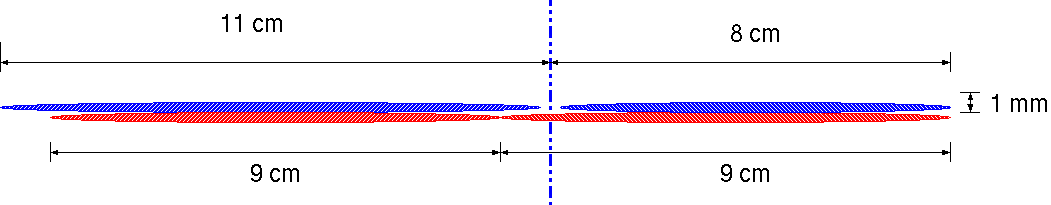}
\caption{A schematic diagram showing the Spin-Light profile at the Ionization Chamber}
\label{fig19}
\end{figure}

\begin{figure}[h!]
\centering
\includegraphics[scale=.25]{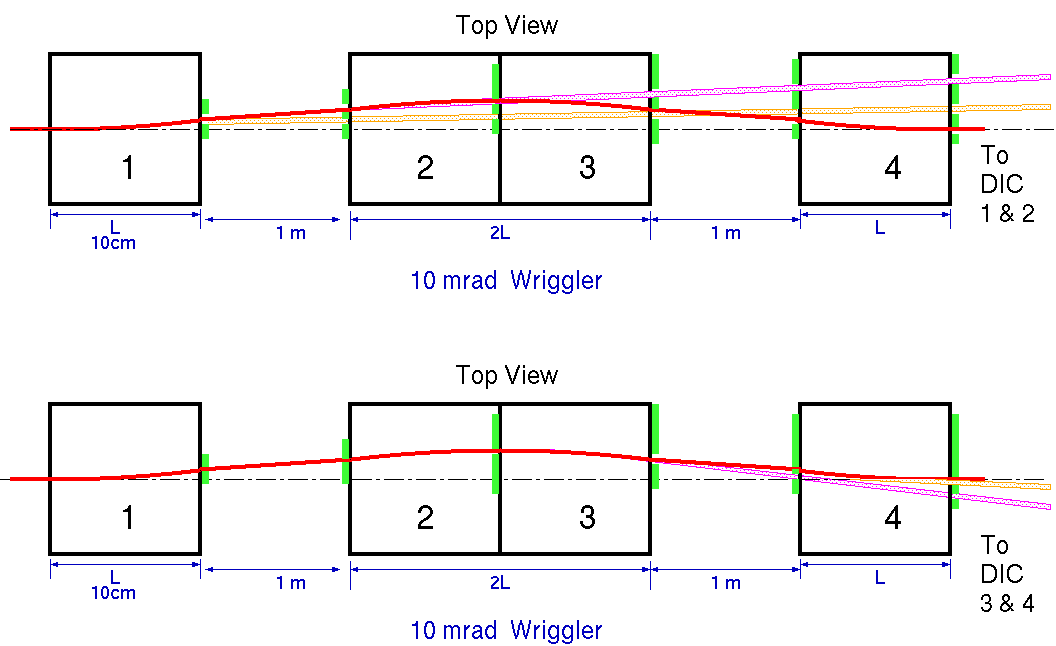}
\caption{A schematic diagram showing the 4 fans of SR that originate at the wiggler magnet system with collimators.}
\label{fig20}
\end{figure}

\begin{figure}[h!]
\centering
\includegraphics[scale=.3]{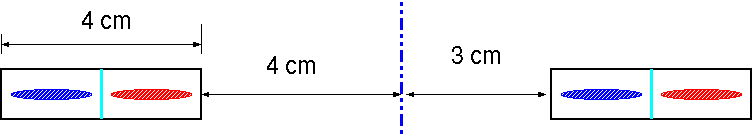}
\caption{A schematic diagram showing the Spin-Light profile at the Ionization Chamber with collimators.}
\label{fig21}
\end{figure}

Figure 18 shows the origin of 4 different fans of SR Light (which contain the spin-light component) that are being created at the wiggler magnet. Corresponding fans of SR light create 4 spots at the ionization chamber which is located $10m$ from the wiggler magnet system. The spots at the ionization chamber as shown in Figure 19, merge with each other and may destroy the spacial asymmetry that contains the polarization information. Therefore collimators may be employed at the face of every dipole to select out a small section of the bigger SR fan as illustrated in Figure 20. After collimation the spots are all uniquely separated and 4 distinct spots can be observed at the ionization chamber (as in Figure 21).

\section{Ionization Chambers}
$\>$ The Spin Light polarimeter detector would consist of a position sensitive ionization chamber to measure the up-down asymmetry in the SR - Light. Such a position sensitive detector that could charecterize X-Ray spectrum has already been developed at the \textit{Advanced Light Source}, Argonne National Laboratory and at \textit{Sprin-8 Light Source}. This uses a split - plane which essentially divides the ionization chamber into $2$ separate chambers but with a common electrode. These have been demonstrated to have a resolution of about $5 \mu m$ $^{[19]}$. Subtracting the currents from the top chamber from the bottom chamber will then give a measure of the asymmetry in the SR-Light. A schematic diagram of the protoype is presented in Figure 23.
\marginpar{$^{[19]}$ K. Sato, J. of Synchrotron Rad., 8, 378 (2001); T. Gog, D. M. Casa and I. Kuzmenko, CMC-
CAT technical report.}

\begin{figure}[h!]
\centering
\includegraphics[scale=.4]{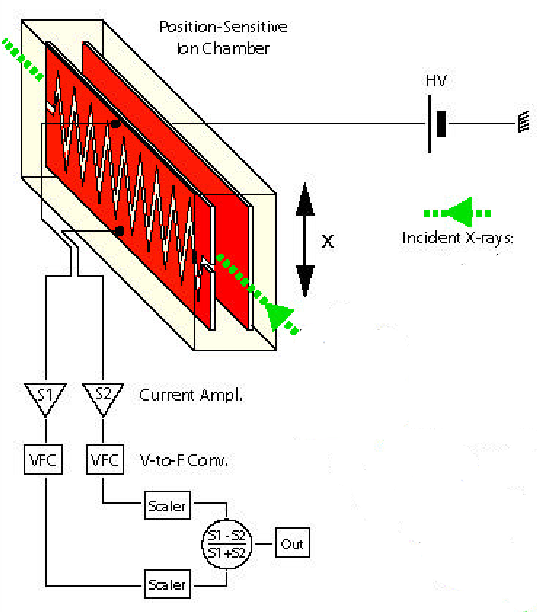}
\caption{A schematic diagram of a prototype split plate Ionization Chamber.}
\label{fig22}
\end{figure}

$\>$ Using such an ionization chamber, one could easily carryout relative polarimetry. A more challenging but possible option would be to have an absolute polarimeter.

\subsection{Relative Polarimetry}
\marginpar{$^{[20]}$ G. Tepper and J. Losee, Nucl. Inst. and Meth. A356, 339 (1995).}

$\>$ A Xenon media split plate would be an ideal differential ionization chamber. Using $Ti$ windows of sufficient size could in principle cut down on low energy X-Rays ($<50KeV$) and $Ti$ has been shown to have a high transparency for hard X-Ray $^{[20]}$. A schematic diagram of the ionization chambers for the Spin-Light polarimeter is presented in Figure 23. The Spin-Light Polarimeter Ionization chamber shall have $2$ compartments into which the $2$ collimated fans of SR Light will enter. On each side of the electron beam is one split - plane ionization chamber and therefore all $4$ fans of SR - Light, produced at the wiggler magnet, are measured at the $2$ ionization chambers. Notice for polarimetry, just one ionization chamber is required. The $2$ separate ICs will provide abundant statistics in a short time.

\begin{figure}[h!]
\centering
\includegraphics[scale=.2]{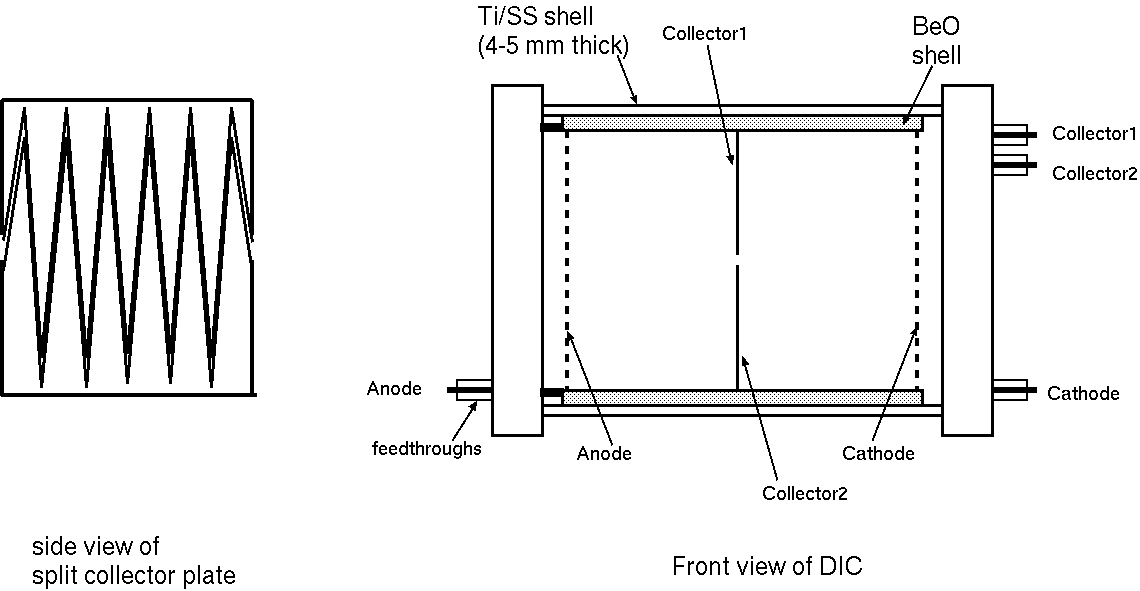}
\caption{A schematic diagram of the Spin - Light Polarimeter Ionization Chamber.}
\label{fig23}
\end{figure}
\begin{figure}[h!]
\centering
\includegraphics[scale=.25]{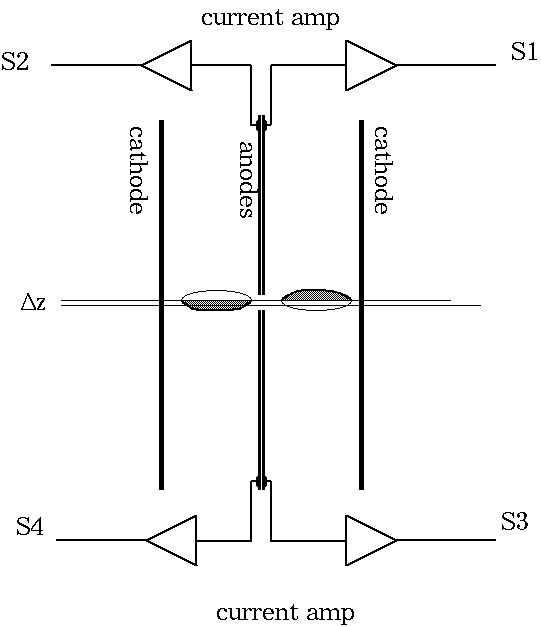}
\caption{A schematic diagram of signal collection configuration.}
\label{fig24}
\end{figure}

The signal which will give us a measure of the spacial asymmetry could be measured by subtracting the currents from the \textsc{UP} and \textsc{DOWN} parts of the chamber after being amplified as shown in Figure 24. The spin - light asymmetry shall be of opposite signs on the \textsc{LEFT} and \text{RIGHT} parts of the chamber, since SR fans from adjacent wiggler dipole (which have opposite polarity) enter one on each side of the chamber (L-R). Beam motion effects are nullified as any motion will have have trend (same sign) on both the LEFT and RIGHT sides of the chambers. Each half (T-B) of the split plane collector measures a current proportional to the difference of photon flux between the $2$ sides and therefore any vertical beam motion effects cancel out to the first order. The two signals indicated in Figure 24 can be quantified. This definitely shows that the vertical beam motion effects will be canceled out to first order.
\begin{equation}
S_1 = N^l_{SR} + N^l_{spin} + \Delta N^l_z - (N^r_{SR} - N^r_{spin} + \Delta N^r_z) = 2N_{spin}
\end{equation}
\begin{equation}
S_2 = N^l_{SR} - N^l_{spin} - \Delta N^l_z - (N^r_{SR} + N^r_{spin} - \Delta N^r_z) = -2N_{spin}
\end{equation}
\marginpar{$^{[21]}$ G. Tepper and J. Losee, Nucl. Inst. and Meth. A356, 339 (1995).}
{\color{blue}\{Where $N^{l(r)}_{SR}$ is the number of SR Photons on the left (right) side of the middle split plate, $N^{l(r)}_{spin}$ is the number of spin-light photons and $\Delta N^{l(r)}_z$ is the difference in number of 
photons introduced by the vertical beam motion\}}.

Hence $S_1 - S_2 = 4N_{spin}$ is a measure of longitudinal polarization and $S_1 + S_2$ will give a measure of transverse polarization. The ability to measure both transverse and longitudinal polarization makes this a powerful polarimetry technique.
The number of photons absorbed in the ionization chamber can be computed by multipling the SR Power equation Eq.(8) with the absorption function (where $\mu$ is the absorption coefficient which is material specific and $t$ is the length of the chamber) $A(\lambda, t) = 1 - e^{- \mu(\lambda).t}$. With the help of values of $\mu$ obtained from NIST database $^{[21]}$, a plot of photons absorbed in a ionization chamber that is $50cm$ in length and held at $1atm$ pressure is shown in Figure 25. The spectra of number of photons absorbed was used to then calculate the detector response which in this case is asymmetry weighted against absorption.
\begin{figure}[h!]
\centering
\includegraphics[scale=.45]{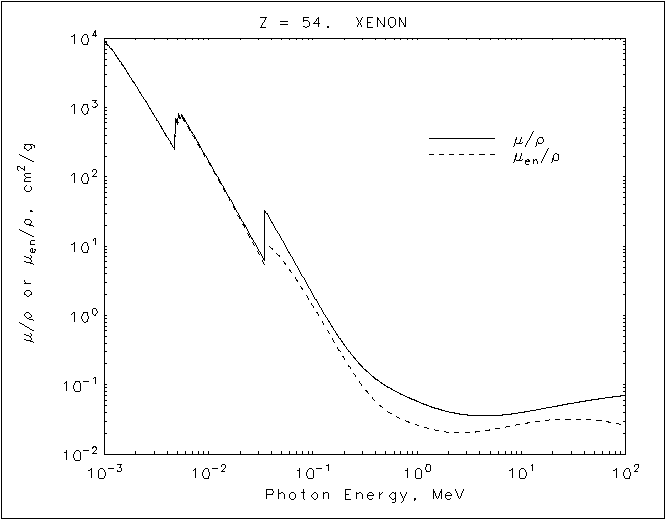}
\caption{NIST plot of dependence of absorption coefficient of Xenon on the photon energy $^{[21]}$.}
\label{fig25}
\end{figure}
\begin{figure}[h!]
\centering
\includegraphics[scale=.45]{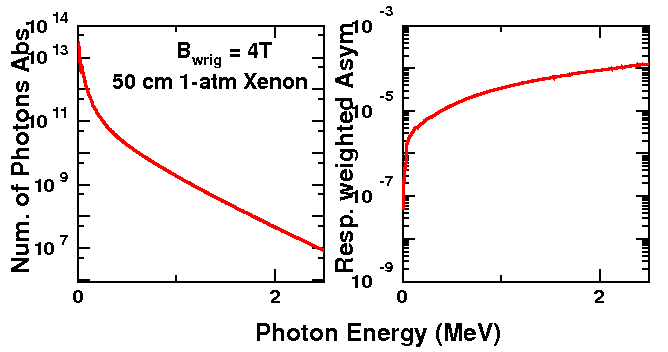}
\caption{Plot of photons absorption spectra for the ionization chamber.}
\label{fig26}
\end{figure}
\begin{figure}[h!]
\centering
\includegraphics[scale=.2]{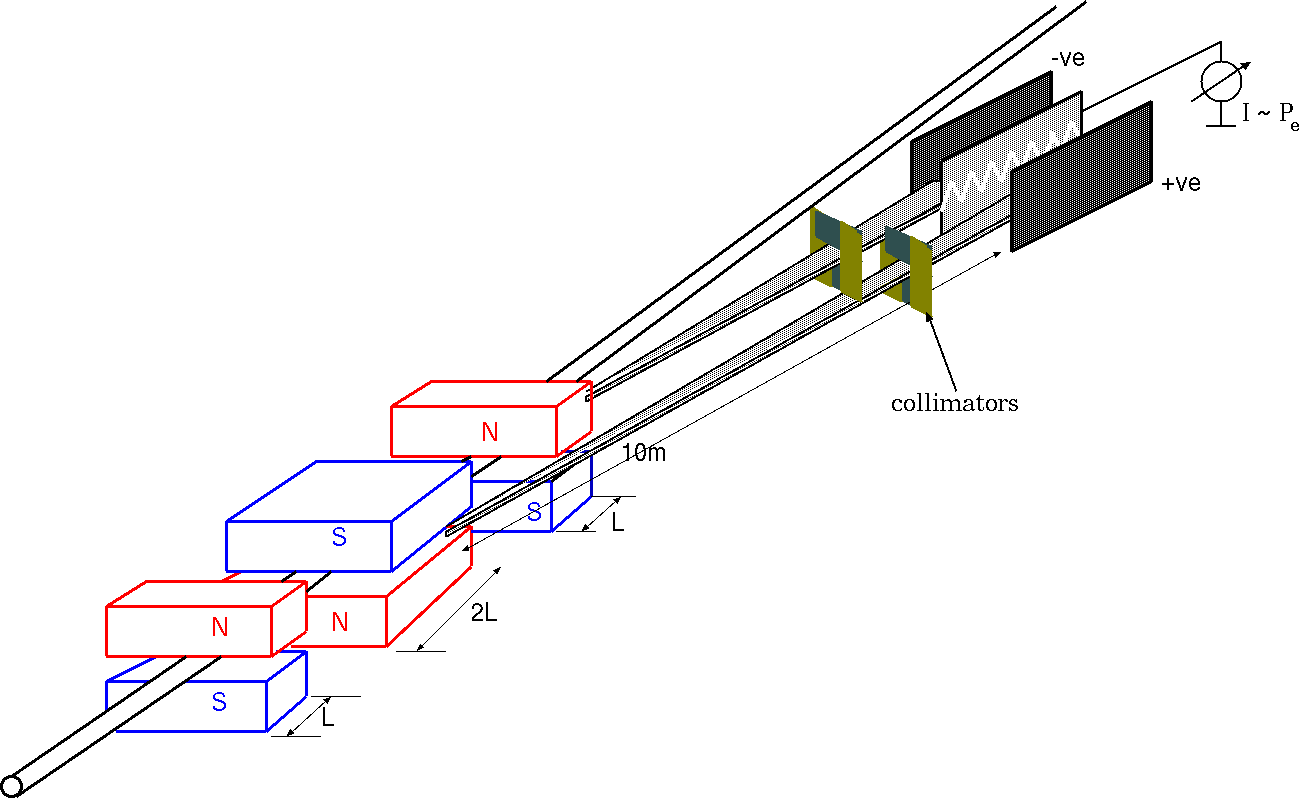}
\caption{Schematic diagram of the entire Differential Spin Light Polarimeter (The only visible difference between the absolute and relative polarimeters in the schematics is the difference in collector plate bias).}
\label{fig27}
\end{figure}
\subsection{Absolute Polarimetry}
\begin{figure}[h!]
\centering
\includegraphics[scale=.2]{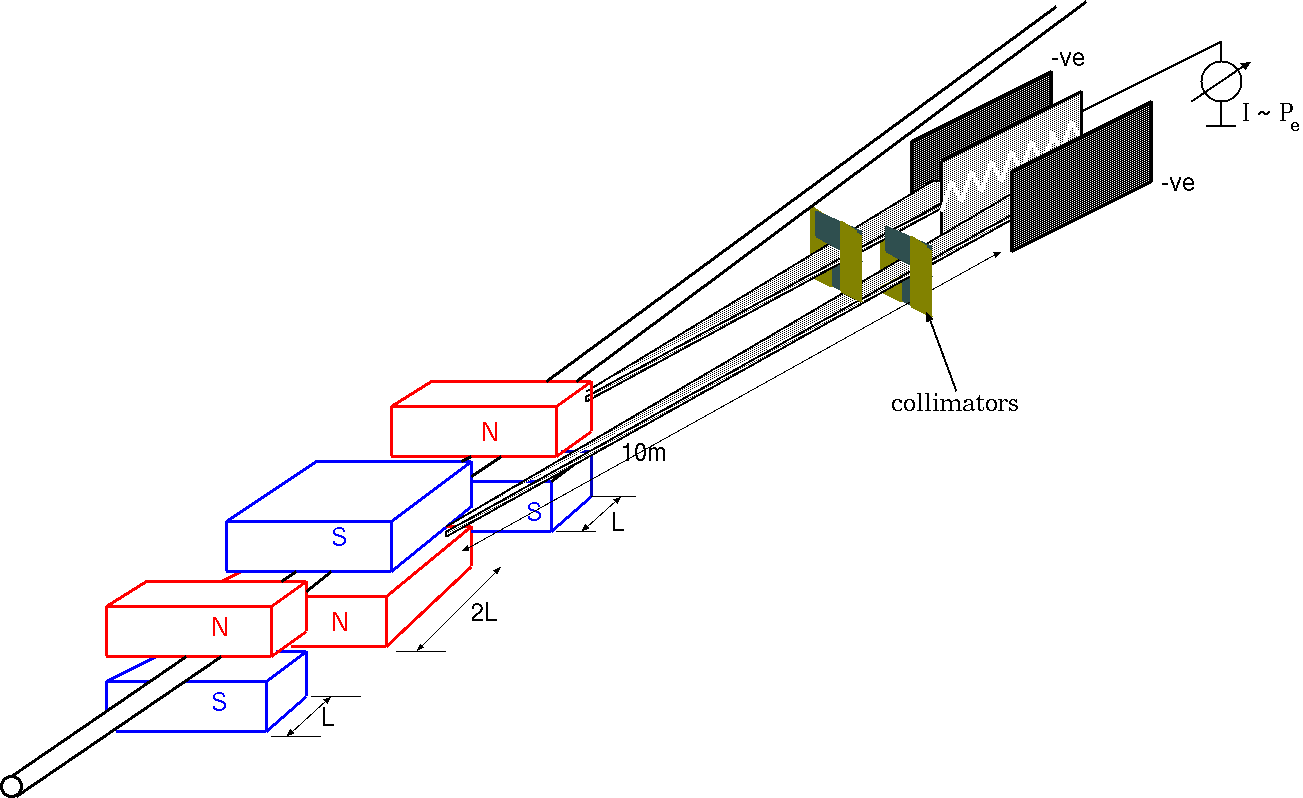}
\caption{Schematic diagram of the entire Absolute Spin Light Polarimeter (The only visible difference between the absolute and relative polarimeters in the schematics is the difference in collector plate bias).}
\label{fig28}
\end{figure}
\marginpar{$^{[22]}$S. Kubota, M. Suzuki and J, Ruan, Phys. Rev. B 21, 2632 (1980).}
A relative polarimeter could be turned into an absolute polarimeter by making a few modifications to the ionization chamber. for absolute polarimeter, a high resolution ionization chamber is required and so the natural choice would be a high pressure Xenon IC. A cylindrical chamber capable of withstanding $50atm$ of pressure could house the IC setup whereas the rest of the structure would remain unchanged from the relative IC with a few additions. The electrodes could be held in place with thin walled $BeO$ ceramic material which would provide uniform electric field and reduce acoustic noise while being transparent to hard XRays $^{[21]}$. This design eliminates the need for field guide rings which require additional feed throughs and internal voltage
dividers. In order to shield against space charge build-up, a wire mesh grid should be placed near the anode which carries a voltage that is intermediate in value to the drift potential (potential between the anode and the cathode). The ratio of the grid field to drift field can be adjusted to maximize the shielding efficiency. The cathodes and the intermediate grids would be build from stainless steel wire mesh to allow the compressed xenon UV scintillation light to be collected by the UV sensitive photomultiplier tubes (PMT). The scintillation signal has a fast component with a decay time of $2.2 ns$ and slow component with a decay time of $27 ns$  $^{[22]}$. The scintillation light can be used during calibration, to provide a time zero reference for ionization position determination and can also be used for background suppression using pulse shape discrimination and for anti-coincidence Compton suppression. This will help improve the energy resolution and hence aid the determination of the sensitive energy range of the chamber (during calibration, when the chamber is operated in charge mode). Similar HPXe chamber (without the split anode) have been successfully operated $^{[21]}$ for over a decade now and are also commercially available. A schematic for such an IC is shown in Figure 29. The readout electronics chain would consist of a pre-amplifier and shaping amplifier unlike the current amplifiers used in the current mode ICs. In addition, one would also have to establish the linearity of such an IC given the high flux of photons making the calibration of the IC very challenging.
\begin{figure}[h]
\centering
\includegraphics[scale=.2]{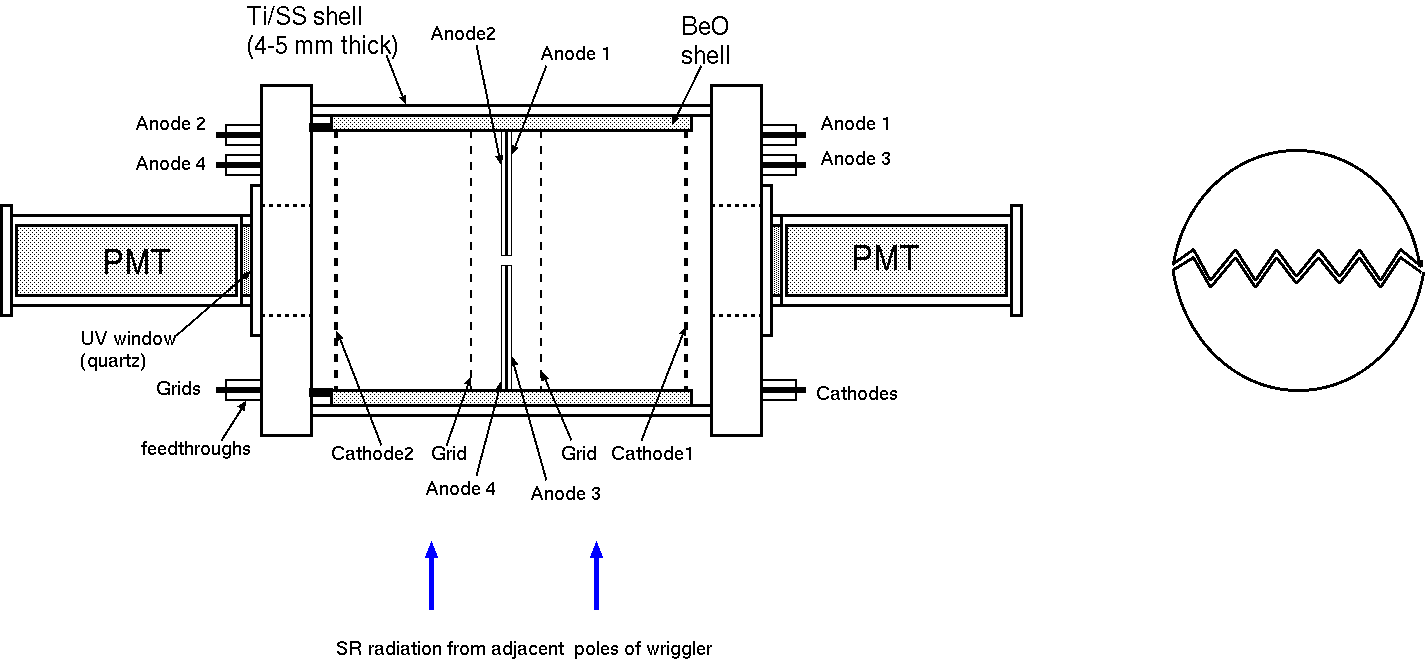}
\caption{A schematic diagram of the Absolute Spin - Light Polarimeter Ionization Chamber.}
\label{fig29}
\end{figure}
The vertical beam motion effects in an absolute IC shall be cancelled to the first order just like in the differential IC. The current signal from each chamber is an integral over the sensitive energy range of the chamber. This energy range convoluted with the detector response function, can be determined by calibrating the chamber at low electron beam currents ($\sim1 nA$), where the photon flux is low enough to operate the chambers in charge mode. The pulse height spectrum from these calibration runs can be used to determine the sensitive energy range and the detector response function. The uncertainty in determining the absolute value of the range of energies integrated (specially the lower bound) is the other major source of uncertainty.

In the case of an absolute IC, $4$ different signals involving each part (TOP/BOTTOM parts of split plane) and (LEFT/RIGHT) parts of the chamber can be tapped for analysis as shown below.
\begin{equation}
S_1 = I_{SR} + I_{spin} + \Delta I_z
\end{equation}
\begin{equation}
S_2 = I_{SR} - I_{spin} + \Delta I_z
\end{equation}
\begin{equation}
S_3 = I_{SR} - I_{spin} - \Delta I_z
\end{equation}
\begin{equation}
S_4 = I_{SR} + I_{spin} - \Delta I_z
\end{equation}
{\color{blue}\{Where $I_{SR}$ is the current due to all SR Photons and $I_{spin}$ is the current due to just the spin- light photons\}}.

The signal $(S_1 + S_2) - (S_3 + S_4)$ should always be zero ideally.  The longitudinal asymmetry in terms of these 4 signals is given by;
\begin{equation}
A^{long} = \frac{I^{long}_{spin}}{I^{long}_{SR}} = \frac{(S_1 - S_2) - (S_3 + S_4)}{(S_1 + S_2) + (S_3 + S_4)}
\end{equation}
and the transverse asymmetry in terms of these 4 signals is given by;
\begin{equation}
A^{trans} = \frac{I^{trans}_{spin}}{I^{trans}_{SR}} = \frac{(S_1 + S_3) - (S_2 + S_4)}{(S_1 + S_3) + (S_2 + S_4)}
\end{equation}
One could in theory come up with many more electrode arrangements.
\subsection{Effects of Extended Beam Size}
\begin{figure}[h!]
\includegraphics[scale=.7]{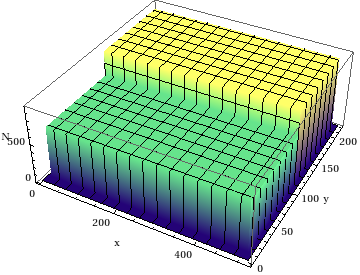}
\caption{Integrated power spectra of SR Light at the IC due to a point-cross section beam - X,Y(10$ \mu m$); N($\times 10^{12}$). (The difference between the profile has been enlarged for clarity)}
\label{fig30}
\end{figure}
\begin{figure}[h!]
\centering
\includegraphics[scale=.7]{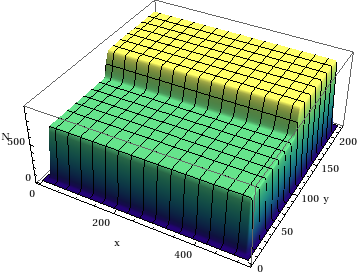}
\caption{Integrated power spectra of SR Light at the IC due to a real beam of size ($R_{beam} = 100\mu m$) - X,Y(10$ \mu m$); N($\times 10^{12}$). (The difference between the profile has been enlarged for clarity)}
\label{fig31}
\end{figure}
$\>$ In \emph{Section 3.2}, the numerical code used a point beam. Therefore the effects of having extended beam size of about $100\mu m$ must be studied. To do this the code located in \emph{Appendix B.3} was used. This code essentially superimposes the SR-Power spectra generated by each of a $10^6$ such point-cross section beams. The million point - cross section beams together would give a circular beam and each of them was weighted with a Gaussian profile in order to make the extended beam a perfect Gaussian beam. The cumulative spectra can be plotted and one can guess that it should have the same structure as the original spectra for the point - cross section beam. This is so because the size of the beam ($R_{beam} = 100 \mu m$) is small compared to the size of the collimated SR - Light spot which is about $1mm$ big. For the beam with a point cross section, the SR - profile is rather 'box' like at the IC. When an extended beam, that is Gaussian profile, is introduced, the SR - profile gets a taper which is Gaussian in nature too. The graphs inn Figure 30 and 31 show the exact 3D  profile correct with position information. 
\chapter{Geant4 Simulations} 

\label{ch:geant4} 
\begin{figure}[h!]
\centering
\includegraphics[scale=.4]{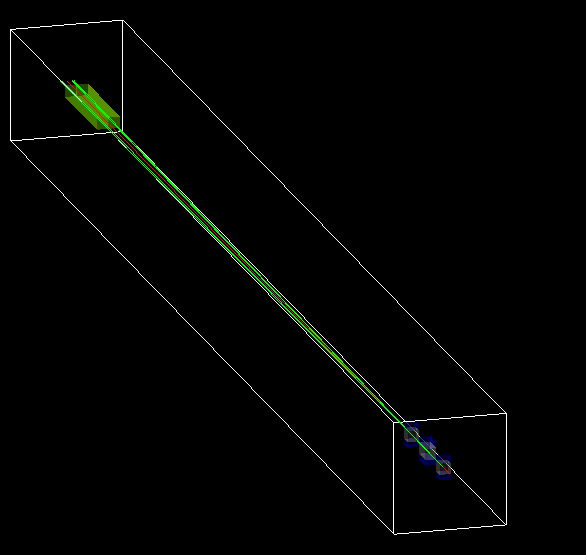}
\caption{GEANT4 visualization of Spin - Light Polarimeter Setup. The electron beam is red in color and the SR Fans are yellow in color.}
\label{fig32}
\end{figure}

\marginpar{$^{[23]}$ F. Herlach, R. McBroom, T. Erber, J. .Murray, and R. Gearhart: Experiments with Megagauss
targets at SLAC, IEEE Trans Nucl Sci, NS 18, 3 (1971) 809-814.}

\marginpar{$^{[24]}$ T. Erber, G. B. Baumgartner, D. White, and H. G. Latal: Megagauss Bremsstrahlung and
Radiation Reaction,Proceedings of High Energy Accelerators (1983), 372-374.}

$\>$ Geant4 has been a widely used took-kit to simulate interactions of particle beams with matter and fields. Originally developed at CERN in Geneva, Switzerland, its development has been decentralized over the last decade. Geant4 does include synchrotron radiation (SR) physics under the standard electromagnetic physics list. 

Geant4 SR process has been rigorously validated at Stanford Linear Accelerator Center in Palo Alto, CA and at Fermi National Accelerator Laboratory in Batavia, IL $^{[23]}$ $^{,[24]}$. It might be important to note that this validation of the SR process is only valid if the beam is made up of electrons in few GeV energy range with SR photons energy spectra peaking at energies much lower than the energy of the beam which means that the electrons passed through magnets with a few tesla field. That is, the Geant4 SR process might not work is the magnets has magnetic fields close to the $B_c$ which is upwards of $100T$. Since the magnetic field in this Spin-Light Polarimeter is near about $4T$, and from Figure 7 (in \emph{Chapter 3}) is can be seen that an 11GeV electron beam (which is one of the proposed electron beam injection energies in both the EIC designs of eRHIC and ELIC), the SR photon energy spectra peaks near about few MeV and therefore using Geant4 to simulate SR photons is justified.  This work demands a Geant4 simulation in order to optimize the distance between the dipoles and to optimize the positioning of the collimators. Figure 32 is a Geant4 visualization of the entire setup. It is important to note that the 2 fans are clearly visible on either side of the beam in the center.

\begin{figure}[h!]
\centering
\includegraphics[scale=.4]{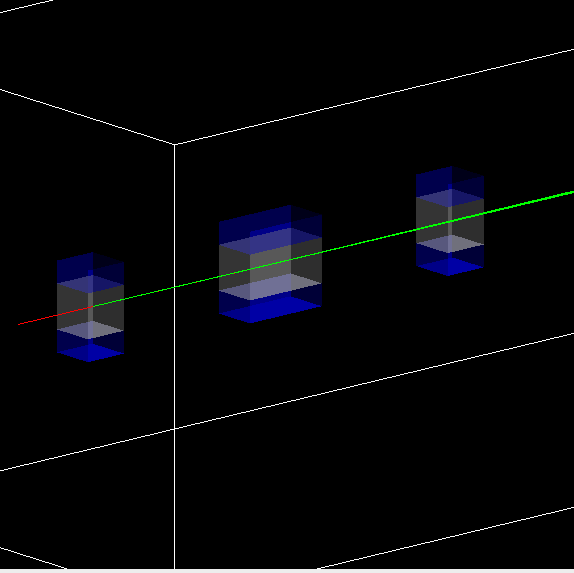}
\caption{GEANT4 visualization of Wiggler Magnet Setup.}
\label{fig33}
\end{figure}
\begin{figure}[h!]
\centering
\includegraphics[scale=.4]{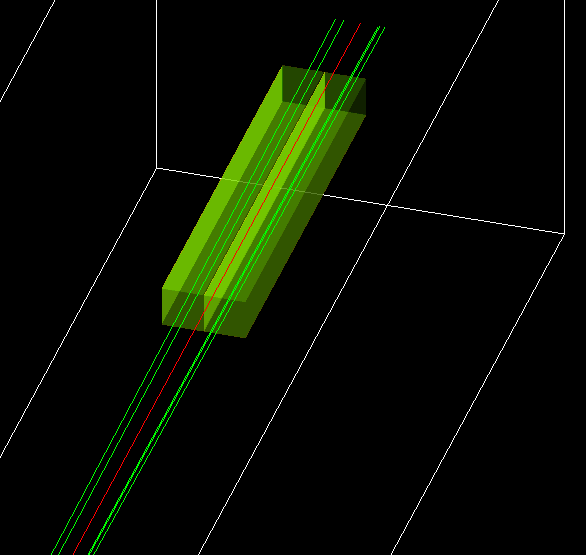}
\caption{GEANT4 visualization of the 2 Ionization Chamber Setups (One on either side of the beam).}
\label{fig34}
\end{figure}

The entire geometry of the setup including the $3$ wiggler magnets, $2$ ionization chambers and, the $5$ collimators were coded into the Geant4. The geometry construction file is available as an appendix (\emph{Appendix B.4}). A visual rendering can be seen in Figure 32. As demonstrated in \emph{Chapter 3}, Figure 19, the collimators select $4$ ($2$ on each side of the electron beam) unique slivers of the bigger SR fan. This can be clearly seen in Figure 34. 

\marginpar{The complete Spin-Light Geant4 Simulation package can be accessed at \url{spinlight.mohanmurthy.com}}

\begin{figure}[h!]
\centering
\includegraphics[scale=.4]{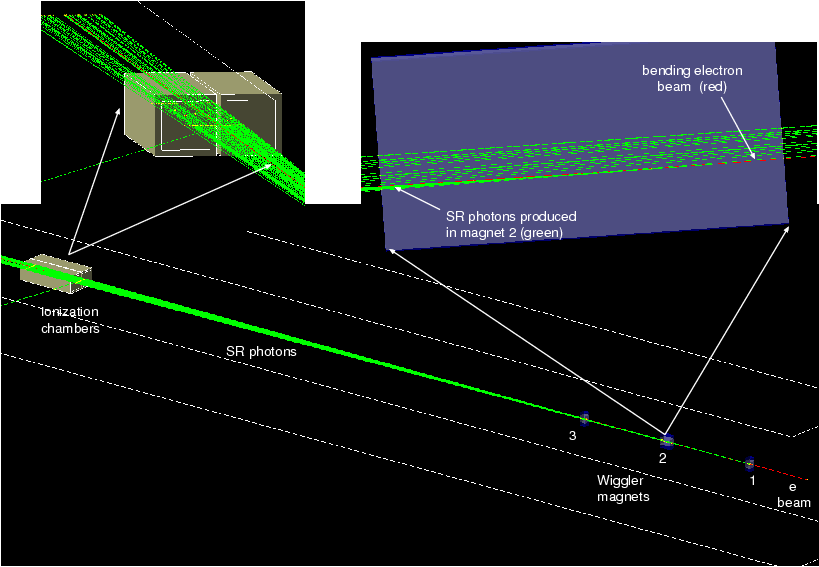}
\caption{SR fans produced by a few 11 GeV electrons.}
\label{fig35}
\end{figure}

It might be fun to spot the electron beam bending in the magnetic field as shown in Figure 35. The code was tested with a few electrons for a $11 GeV$ beam. Some simple observations that could be made straight away without measuring anything would be;
\begin{itemize}
\item The electron beam returns to its original linear trajectory (as along the initial electron gun firing vector). This means that the magnets are aptly separated with appropriate fields.
\item The red colored electrons produce yellow colored photons in conical fans.
\end{itemize}

\section{Spin Light in Geant4 and Asymmetry}

$ \> $ Even though the integrated energy spectrum (over all angles which includes above and below the electron orbit in the wiggler magnets) is validated, the angular dependence of SR light is not in Geant4. Therefore one would have to add the angular dependence in Equation 4 (\emph{Chapter 1}) to the Geant4 SR process in the standard electromagnetic list. Alternatively, one could modify their application created using standard Geant4.9.6 source code and add the SR angular dependence into their stacking action.  
Our Spin-Light simulation application implements the Spin-Light by manipulating the spacial asymmetry according to  \emph{Section 3.1}. The asymmetry can be linear polynomial parameterized as:

\begin{multline} 
A(\textbf{E}) = (4.5 \times 10^{-7}) + (5.7 \times 10^{-5} {\textbf E}) - (3.0 \times 10^{-5}{\textbf E^2}) + (1.3 \times 10^{-5} {\textbf E^3}) \\ - (3.3 \times 10^{-6} {\textbf E^4}) + (3.2 \times 10^{-7} {\textbf E^5}) + \Theta({\textbf E^6}) ... 
\end{multline}
{\color{blue}\{\textit{A} is the asymmetry and \textit{E} is the energy of the SR photon\}}

At the event stacking level, the SR photons are killed with a probability equal to asymmetry calculated by Equation 25. This creates an up-down asymmetry in the SR cones which is vital for successful simulation of a Spin-Light polarimeter. 

\begin{figure}[h!]
\centering
\includegraphics[scale=.5]{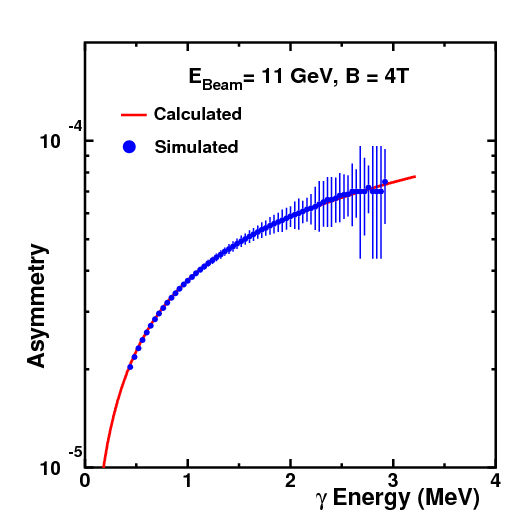}
\caption{Geant4 Asymmetry compared with physics asymmetry}
\label{fig36}
\end{figure}

As seen in Figure 36, the simulated asymmetry closely matches the physics asymmetry curve within $1\%$. It is important to note that the parameterization shown in only good in relevant range of photon energies ($0-3.5GeV$). Anyone who would like to use this parameterization beyond this relevant range would have to fall back on the numerical code in \emph{Appendix B.1}.

A histogram of SR photon energies is then constructed. By dividing the difference of number of photons with positive momentum and number of photons with negative momentum with the total number of SR photons, a spin-light spectrum was made as shown in Figure 37. But before plotting the histograms were scaled up since some photon tracks were killed in the stacking action. This should be acceptable as the asymmetry is usually very small and so the scale factor is very close to $1$ ($\sim 1.0001$). In Figure 37, the power spectra seems to deviate at low energies (close to about $0.5 GeV$. This is the case as the SR photon characteristics plotted in Figure 37 are characteristics of SR photons as detected at the IC and for this measurement, a $0.3mm$ solid lead window was used to eliminate low energy X-Rays below $0.5 GeV$. Eliminating X-Rays below $0.5 GeV$ is particularly important as most of the SR photons are low energy photons, but the asymmetry is low at low energies contributing to very small numbers of spin-light photons. In essence, the low energy photons are junk! Ones it is made sure that the energy spectra matches the physics prediction very closely, the corresponding power spectra is easy to calculate and plot as in Figure 38.

\begin{figure}[h!]
\centering
\includegraphics[scale=.5]{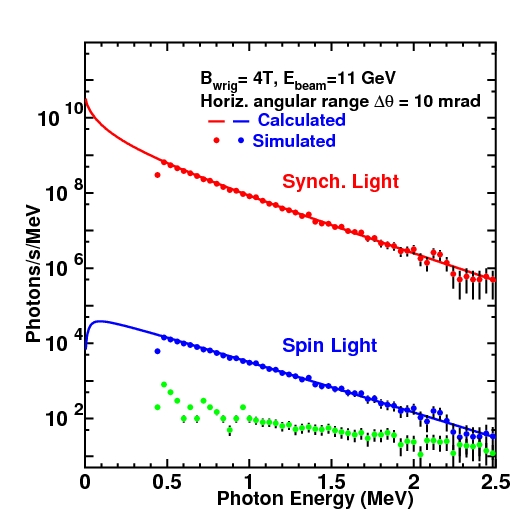}
\caption{Geant4 SR and SL spectra as compared with physics SR and SL spectra}.
\label{fig37}
\end{figure}

\begin{figure}[h!]
\centering
\includegraphics[scale=.7]{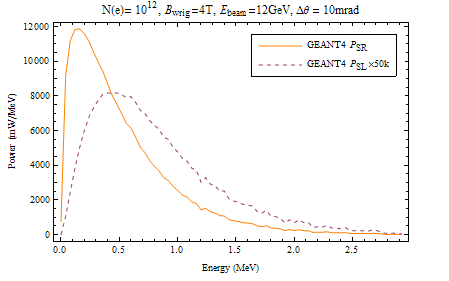}
\caption{Geant4 SR and SL power spectra}.
\label{fig38}
\end{figure}

\section{Synchrotron Light Profile at the Ionization Chambers and Collimation}

$\>$ The collimators are vital to separate the fans from each magnet so that the spacial asymmetry is intact as predicted by the simple physics model that was developed in \emph{Chapters  1 \& 2}. Without the collimators the SR light profiles would all mix and might not necessarily help linear polarimetry (linear to the polarization of electrons in the beam). Therefore, the $4$ slivers selected by $5$ collimators must not mix and the ionization events created by these SR photons should be clearly resolvable in space at the ionization chambers. With the collimator opening width of about $100\mu m$, the $4$ SR slivers are clearly resolved as shown in Figure 39. 

\begin{figure}[h!]
\centering
\includegraphics[scale=.35]{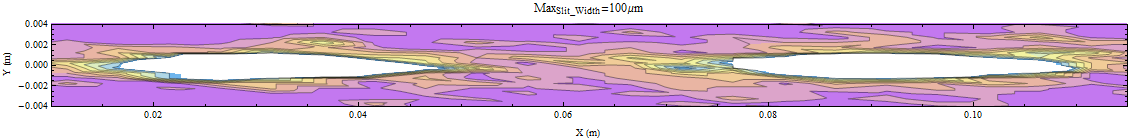}
\caption{A contour plot of the number of events generated by the $2$ slivers of the SR fan in one of the ionization chambers. The white regions are regions of high event density and the blue regions have a low event density.}
\label{fig39}
\end{figure}

\begin{figure}[h!]
\centering
\includegraphics[scale=.35]{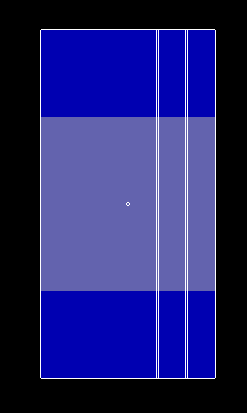}
\caption{Position of collimators on the face of magnet 2 facing magnet 1.}
\label{fig39}
\end{figure} 

With the help of Figure 20 (\emph{Section 3.4}), it can be seen that photons in each [of the 2] spot in the Figure 39 comes from a unique dipole magnet (numbered $1$ through $4$, the central dipole consists of $2$ "parts" as it is twice as long as the other two). Even though each spot is only suppose to contain photons from one specific [dipole magnet] part, a source of error is introduced here since this is not always the case. For example consider the IC on the left hand side of the beam which has spots from dipole magnet part \# $1$ \& $2$. In this case, the left hand side spot in Figure 39 should contain photons from dipole magnet part $1$ and the right hand side spot must contain photons from dipole magnet part $2$. But, some photons from dipole magnet part $1$ seep into the spot on the right hand side. If the percentage of these photons is $\sim 0.1\%$ we consider the spots resolved. The size of the collimators is an important factor which determines if these spots are completely resolved. With a collimator slit width of about $100 \mu m$ the spots are completely resolved. Another source of error here could be from bremsstrahlung events from the photons impinging on the collimators. These bremsstrahlung photons contribute to no spin-light events and therefore are junk events. Fortunately these events comprise of about only $\sim 0.1\%$. 
\chapter{Summary} 

\label{ch:sim} 

\section{Systematics}
$\>$ If the ionization chambers are used in differential mode and have split anodes, the false asymmetries will
cancel to first order. Moreover, since the signal used is a differential signal the size of the background
must be small compared to the signal. A full simulation is needed to study the background and the
asymmetry associated with the background. In the experiment the background can be determined
by monitoring the difference in the signal from the chambers with the wriggler magnets turned on
and off. The other major source of systematic uncertainty is the lower bound of the integration
window used to generate the IC signals . The absolute value of the spin light asymmetry depends
on the absolute value of the energy window over which the IC signals are integrated. It is especially
sensitive to the lower bound because of the steep fall of the SR intensity with energy. However
given the excellent energy resolution that has been demonstrated for HPXe ionization chambers,
one should be able to calibrate the chamber and determine the response function and the lower
bound of the chamber to better than $2\%$. A preliminary table of estimated systematic 
uncertainties is shown in table below.
\begin{table}[h]
\centering
\begin{tabular}{l|l|c}
Source & Uncertainity & $\delta A/A$\\
\hline
Dark Current & ~$pA$ & $<0.01\%$\\
Intensity Fluctuations & $\Delta N \times 10^{-3}$ & $<0.10 \%$ \\
Beam Energy & $1.0 \times  $ & $<0.05\%$ \\
Density of gas in IC & Relative uncertainties & $<0.05\%$\\
Length of Chamber & Can be corrected & -\\
Band - width of X - Rays & $2\%$ (for only absolute polarimetry) &  ~$1.20\%$ \\
Background related Dilutions & To be determined if known to $0.5\%$ &  $<0.50\%$ \\
Other dilutions & Cancel to First Order & $<0.50\%$ \\
\hline
Total & Relative Polarimetry & $<0.68\%$  \\
 & Absolute Polarimetry & $<1.88\%$\\
\hline
\end{tabular}
\caption{Systematic uncertainties}
\label{tab:table1}
\end{table}

\section{Conclusion and Future Work}
$\>$ Spin light based polarimetry was demonstrated over 30 years ago, but has been ignored since then.
The figure of merit for such a polarimeter increases with electron beam energy and the strength of
magnetic field used. The $11 GeV$ beam at JLab is well suited for testing a spin light polarimetry and such a
polarimeter would help achieve the $~0.5 \%$ polarimetry desired by experiments envisioned for the
EIC era. A 3 pole wriggler with a field strength of $4 T$ and a pole length of $10 cm$ would be
adequate for such a polarimeter. A dual position sensitive ionization chambers with split anode
plates is ideally suited as the X-ray detector for such a polarimeter. The differential detector design
would help reduce systematic uncertainties. Locating a reasonable piece of beam-line real estate is
however very challenging.

\marginpar{$^{[25]}$ V. N. Litvinenko, Gatling Gun:  High Average Polarized Current Injector for eRHIC, EIC \@ BNL Whitepapers (2012)}

Given that the eRHIC design of EIC involves using a Gatalin gun $^{[25]}$, the recovery time of the spin-light IC will need to be less than a second if every bunch of electron beam is to be measured for polarization. So the next thing to do would be create a survey of all experiments beings proposed and their corresponding polarimetry requirements both in terms of precision of polarimetry required and the rate of measurement. Finally it still remains a challenge to secure ample grants in excess of quarter million dollars over the next few years in order to construct a prototype of a Spin-Light Polarimeter for demonstration purposes. VEPP storage ring in Novosibirsk could be a place to test such a prototype. 


\appendix



\chapter{Lists and Permissions}

\section{List of Figures}
\begin{center}
\begin{longtable}{r|l|l}
Fig. & Sec. & Caption (in short)\\
\hline
i & For. & {\bf eRHIC} @ Brookhaven National Laboratory.\\
ii & For. & {\bf ELIC} @ Jefferson National Accelerator Laboratory.\\
1 & 1.1 & Angular distribution of synchrotron radiation.\\
2 & 1.2 & Magnetic snake used at the VEPP-4. \\
3 & 1.2 & VEPP-4 Polarization Experiment.\\
4 & 2.1 & 3-Pole Wiggler.\\
5 & 2.2 & 4 SR fans created by the wiggler magnets.\\
6 & 2.3 & $Cs^{137}$ events in a Xenon IC.\\
7. & 3.1 & SR-SL spectra, Asymmetry \& Time requirement. \\
8. & 3.1 & SL, Asymmetry  spectra dependence on dipole field.\\
9. & 3.1 & SR, Asymmetry  spectra dependence on dipole field.\\
10. & 3.1 & Pole Length \& Time requirement with dipole field.\\
11. & 3.2 & Schematic of half a dipole.\\
12. & 3.2 & Field map at the center of a dipole.\\
13. & 3.2 & Magnetic field at the center of a dipole.\\
14. & 3.2 & Field map at the face of a dipole.\\
15. & 3.2 & Magnetic field at the face of a dipole.\\
16. & 3.3 & SR, SL spectra with realistic magnetic field.\\
17. & 3.3 & Asymmetry with realistic magnetic field.\\
18. & 3.4 & Schematic of 4 X SR Fans from wiggler.\\
19. & 3.4 & Schematic of the 4 X spots at IC.\\
20. & 3.4 & Slivers of IC selected by the Collimator.\\
21. & 3.4 & Schematic of 4 X spots after Collimation.\\
22. & 3.4 & Schematic of Split plane IC.\\
23. & 3.4 & Schematic of Split plane IC, side view.\\
24. & 3.4 & IC Signal Collection.\\
25. & 3.4 & Xenon Absorption coefficient spectra.\\
26. & 3.4 & SR Absorption and Asymmetry spectra.\\
27. & 3.4 & Schematic of differential polarimeter.\\
28. & 3.4 & Schematic of absolute polarimeter.\\
29. & 3.4 & Schematic of Absolute Split plane IC, side view.\\
30. & 3.4 & Integrated SR Spectrum at IC with point beam.\\
31. & 3.3 & Integrated SR Spectrum at IC with Gaussian beam.\\
32. & 4.0 & Geant4 visualization of Spin-Light Polarimeter.\\
33. & 4.0 & Geant4 visualization of wiggler magnets.\\
34. & 4.0 & Geant4 visualization of ICs.\\
35. & 4.1 & Few event visualization in the Polarimeter.\\
36. & 4.1 & Geant4 Asymmetry Spectra.\\
37. & 4.1 & Geant4 SR-SL Spectra.\\
38. & 4.2 & Geant4 SR-SL Power Spectra.\\
39. & 4.2 & Collimated SR Spots at IC.\\
40. & 4.2 & Collimators on magnet 2 back face.\\
\hline

\caption{List of Figures}
\label{tab:table2}
\end{longtable}
\end{center}

\section{List of Referenced Publications}
\begin{center}
\begin{longtable}{r|l}
[\#] & Publication\\
\hline
i & eRHIC, \url{www.bnl.gov/cad/eRhic/}, 2011.\\
ii & ELIC @ CEBAF, \url{casa.jlab.org/research/elic/elic.shtml}, 2011.\\
1 & D.Dutta et. al, Proposal to the EIC R \& D: \url{spinlight.mohanmurthy.com}.\\
2 & D. D. Ivanenko et. al.,16, 370 (1946)\\
2 & J. Schwinger, Phys. Rev. 75, 1912 (1947)\\
3 & I. M. Ternov et. al., Univ. Ser. Fiz. Astr. 24, 69 (1983) \\
3 & V. A. Bordovitsyn et. al., Nucl. Inst. and Meth. B266, 3708 (2008)\\
4 & A. A. Sokolov et. al., JETF 23, 632 (1952). \\
5 & A. A. Sokolov et. al., Rad. from Rel. $e^-$, A.I.P. Transl. Sr. (1986).\\
5 &  I. M. Ternov, Physics - Uspekhi 38, 409 (1995).\\
5 & S. A. Belomesthnykh et al., Nucl. Inst. and Meth. 227, 173 (1984).\\
6 & K. Sato, J. of Synchrotron Rad., 8, 378 (2001).\\
7 & J. Le Duff et. al., Rapport Technique, 4-73 (1973).\\
8 & A. E. Bolotnikov et. al., Nucl. Inst. and Meth. A396, 360 (1997).\\
9 & T. Doke, Portugal Phys. 12, 9 (1981).\\
10 & V. V. Dmitrenko et al., Sov. Phys.-tech. Phys. 28, 1440 (1983).\\
10 & A. E. Bolotnikov et al., Sov. Phys.-Tech. Phys. 33, 449 (1988).\\
11 & C. Levin et al., Nucl. Inst. and Meth. A332, 206 (1993). \\
12 & G. Tepper et. al., Nucl. Inst. and Meth. A356, 339 (1995).\\
13 & A. E. Bolotnikov et. al., Nucl. Inst. and Meth. A383, 619 (1996).\\
14 & G. Tepper at. al., Nucl. Inst. and Meth. A368, 862 (1996).\\
15 & Proportional Technologies Inc., \url{www.proportionaltech.com}.\\
16 & B. Norum, CEBAF Technical note, TN-0019 (1985).\\
17 & M. Sands, SLAC Technical note, SLAC-121 (1970).\\
18 & EM Solver, \url{laacg1.lanl.gov/laacg/services/sfu\_04\_04\_03.phtml},2007.\\
19 & K. Sato, J. of Synchrotron Rad., 8, 378 (2001).\\
19 & T. Gog et. al., CMC-CAT technical report.\\
20 & G. Tepper and J. Losee, Nucl. Inst. and Meth. A356, 339 (1995).\\
21 & G. Tepper and J. Losee, Nucl. Inst. and Meth. A356, 339 (1995).\\
22 & S. Kubota, M. Suzuki and J, Ruan, Phys. Rev. B 21, 2632 (1980).\\
23 & F. Herlach et. al., IEEE Trans Nucl Sci, NS 18, 3 (1971) 809-814.\\
24 & T. Erber et. al.,Proc. of High Energy Accelerators (1983), 372-374.\\
25 & V. N. Litvinenko, Gatling Gun for eRHIC \@ BNL Whitepapers (2012).\\
\hline

\caption{List of Reference Publications}
\label{tab:table3}
\end{longtable}
\end{center}

\section{License}

\paragraph{GNU General Public License:} All the programs are free software; you can redistribute it and/or modify it under the terms of the \textsmaller{GNU} General Public License as published by the Free Software Foundation; either version 2 of the License, or (at your option) any later version.

This program is distributed in the hope that it will be useful, but \emph{without any warranty}; without even the implied warranty of \emph{merchantability} or \emph{fitness for a particular purpose}. See the \textsmaller{GNU} General Public License for more details. 

\chapter{Essential Codes}

\section{Numerical Integration of the SR - Power Law}
\lstset{language=Fortran, caption=Fortran 77 Code, label=numint}
\begin{lstlisting}
C
C      PROGRAM Ngamma.f (WRITTEN BY D. Dutta 1/7/2010)
C      and
C      Ngammaspectra.f by Prajwal Mohanmurthy (Sept 2011), 
C      prajwal@jlab.org, Mississippi State University
C
C      To include effects from the Real Magnetic field 
C      with non-zero gradient taper.
C
C      This program calculates the total number (and 
C      difference in number above and below the orbital
C      plane) of Synchrotron photons emitted by longitudinally
C      polarized electrons over a horizontal angular range of 
C      dTheta and verticle angular range of;
C                         +/-alpha = +/-gamma*Psi
C      where gamma is the Lorentz boost, i.e.;
C                         +/-alpha approx = +/-1 
C      when traversing a 3 pole wriggler magnet with a field 
C      stength of Bwg tesla and a pole length of Lwg. 
C
C       IMPLICIT DOUBLE PRECISION (A-H,O-Z)
C      
       IMPLICIT real*8(A-H,O-Z)
       external fn01
       external fn02
       external gn01
       external gn02
       real*8 xheeng(48),xhemu(48),couB(15)
       real*8 couxn,coudn

       parameter (xMe=0.510998902) !electron mass in MeV/c^2
       parameter (GeV2MeV=1000.0) 
       parameter (hbarc=197.3269602) ! MeV*fm
       parameter (xmuB=5.788381749E-11) ! Bohr magnetron MeV/T
       parameter (c=299792458) ! m/s
       parameter (pi=3.141592654)
       parameter (qe=1.602176462E-19) ! coulomb
       parameter (n1=10) ! number of times the integration alg is compounded
       parameter (n2=10)

       Common gamma,y

c       write(6,*)'Enter Ebeam (GeV) and current (micro A)'
c       read(5,*)Ebeam,xIe
c       write(6,*)'Wriggler B-field (T) and Pole length (m)' 
c       read(5,*)Bwg,xLwg
       do i = 1,15
         couB(i)=4.55
       enddo
       couB(1)=4.11
       couB(2)=4.38
       couB(14)=4.38
       couB(15)=4.11
       open(unit=10,file="spinlight_gydep4.dat",status="unknown")
       open(unit=11,file="xenon.dat",status="old")
       do i =1,48
         read(11,*)xheeng(i),xhemu(i)
       enddo
       close(11)
       ymin = 0.01 ! min fractional photon freq (W/Wc)
       ymax=0.02 !initialize
c       Ebeam=4.0+(i-1)*1.0 ! e-beam enengy GeV
       Ebeam=11.0 ! e-beam enengy GeV
       xIe=100.0  ! e-beam current micro A
       do i=1,1000
        do j=1,15
         couxn=0.0
         coudn=0.0  
         Bwg=couB(j)    ! B-field in T
c       Bwg=1.0+(i-1)*1.0
c       xLwg=0.066  ! pole length in m
c       write(6,*)'Ebeam =',Ebeam, 'GeV'
         gamma = Ebeam*GeV2MeV/xMe  ! Lorentz boost = E/(Me*c^2)
         R_bend = gamma*hbarc*1.0E-15/(2.*xmuB*Bwg) !bending radius in m
         Omega_o = c/R_bend  ! betatron freq.
         Omega_c = 1.5*gamma**3*Omega_o ! central photon frequency
         E_cent=(Omega_c*hbarc*1.0E-15/c)*1000. ! central photon energy in keV
         xlambda_c=2.*pi*c/Omega_c
         ymax = 0.02+(i-1)*0.01 ! max fractional photon freq (W/Wc)
         y_cent= (ymin+ymax)/2.
         E_min = (ymin*Omega_c*hbarc*1.0E-15/c)*1000. ! min photon energy in keV 
         E_max = (ymax*Omega_c*hbarc*1.0E-15/c)*1000. ! max photon energy in keV
         E_cent =(y_cent*Omega_c*hbarc*1.0E-15/c) ! photon energy in MeV 
         xLwg = 0.0135 ! pole length in m
         dTheta = xLwg/R_bend ! horizontal angular range
         ij=2
         ik=1
         ift=1
         do ii=1,48
           if(E_cent.lt.xheeng(ii).and.ift.eq.1)then
             ij=ii
             ik=ii-1
             ift=ift+1
           endif
         enddo
         xrayabs=xhemu(ik)+
     > (((E_cent-xheeng(ik))/(xheeng(ij)-xheeng(ik)))*
     >                          (xhemu(ij)-xhemu(ik)))
c       absconst=0.023*(2./4.)*(2.*xlambda_c*1.0E+10/y_cent)**2.78
c       if (absconst.lt.0.1) absconst=0.1
c       write(6,*)E_cent,xrayabs
         absconst=xrayabs*(5.9/1000.) 
         Amut=(1.0-exp(-absconst**0.5))
c       write(6,*)'Wriggler B-field and pole lengths =',Bwg,xLwg
c       write(6,*)'boost, central freq and bend radius=',gamma,Omega_c,
c     1            R_bend
c       write(6,*)'vertical ang. range, min and max photon energy=',
c     1 dTheta, E_min, E_max,xlambda_c,absconst,y_cent,Amut,E_cent
c       Psi_min = -asin(1./gamma) ! min vertical angle
c       Psi_max = asin(1./gamma) ! max vertical angle
c       alpha_min = gamma*Psi_min ! boosted min vertical angle
c       alpha_max = gamma*Psi_max ! boosted max vertical angle
c       z_min = (ymin/2.)*(1.+alpha_min**2.)**1.5 ! z=(y/2)*(1+alpha)^1.5
c       z_max = (ymax/2.)*(1.+alpha_max**2.)**1.5
c       write(6,*)alpha_min,alpha_max,z_min,z_max
         xNe=xIe*1.0E-06/qe  ! # of electrons
         xi=1.5*gamma**2*hbarc*1.0E-15/xMe/R_bend ! crit parameter
         xHbyHo=gamma*hbarc*1.0E-15/xMe/R_bend
         tau=(8.*sqrt(3.)/15.)*(hbarc*1.0E-15/xMe/c) ! use hbarc/qe^2 =137
     1        *(1./xHbyHo)**3*(1./gamma**2.)*137.0
         sflip=hbarc*(xLwg/c)*(1.+8.*sqrt(3.)/15.)*0.5*1./tau
c       write(6,*)'sflip probability',sflip
         const1=3.*xNe*gamma*dTheta/(4.*pi**2.*137.)
         const2=4.*const1*xi
         call p3pgs(ymin,ymax,n1,fn01,gn01,vint1) ! integrations for N
         call p3pgs(ymin,ymax,n2,fn02,gn02,vint2) ! integraton for Delta N
         xn=const1*vint1
         dn=const2*vint2
         xa=(dn/xn)*sqrt(2.*xn)
         xpow=xn*E_cent*1.6E-19*1.0E+6! power released in W
         dpow1=dn*E_cent*1.6E-19*1.0E+10 ! power of spin light in W
c       write(6,*)'edep',xdpow !Ebeam,gamma,dTheta,gamma*dTheta,const1,vint1
c       write(6,*)'photon energy, #of photons, up/dn diff, assym, 
c     1 analyzing pwr, photons abs'
c       write(6,15)(E_min+E_max)/2000.,xn,dn,dn/xn,xa,xpow,Amut*xn
c	y_cent = E/E_max (do not add), E_cent: bi central energy(do not add), xn: tot. num. events (add),dn:Events from Spin light(add), dn/xn = ratio, xa= events above - events below/ sum, xpow: integrated per bin, Amut*xn: signal size  
         couxn = couxn + xn
	 coudn = coudn + dn
        enddo
        xn = couxn
        dn = coudn
        write(10,15)y_cent,E_cent,xn,dn,dn/xn,xa,xpow,Amut*xn
        ymin=ymax
       enddo
       close (10)
 15    format(1x,f6.3,1x,f6.3,1x,e15.3,1x,e15.3,1x,e15.3,1x,e15.3,1x,
     1        e15.3,1x,e15.3)
       END

        SUBROUTINE IKV(V,X,VM,BI,DI,BK,DK)
C
C       =======================================================
C       Purpose: Compute modified Bessel functions Iv(x) and
C                Kv(x), and their derivatives
C       Input :  x --- Argument ( x > 0 )
C                v --- Order of Iv(x) and Kv(x)
C                      ( v = n+v0, n = 0,1,2,..., 0 < v0 < 1 )
C       Output:  BI(n) --- In+v0(x)
C                DI(n) --- In+v0'(x)
C                BK(n) --- Kn+v0(x)
C                DK(n) --- Kn+v0'(x)
C                VM --- Highest order computed
C       Routines called:
C            (1) GAMMA for computing the gamma function
C            (2) MSTA1 and MSTA2 to compute the starting
C                point for backward recurrence
C       =======================================================
C
        IMPLICIT DOUBLE PRECISION (A-H,O-Z)
        DIMENSION BI(0:*),DI(0:*),BK(0:*),DK(0:*)
        PI=3.141592653589793D0
        X2=X*X
        N=INT(V)
        V0=V-N
        IF (N.EQ.0) N=1
        IF (X.LT.1.0D-100) THEN
           DO 10 K=0,N
              BI(K)=0.0D0
              DI(K)=0.0D0
              BK(K)=-1.0D+300
10            DK(K)=1.0D+300
           IF (V.EQ.0.0) THEN
              BI(0)=1.0D0
              DI(1)=0.5D0
           ENDIF
           VM=V
           RETURN
        ENDIF
        PIV=PI*V0
        VT=4.0D0*V0*V0
        IF (V0.EQ.0.0D0) THEN
           A1=1.0D0
        ELSE
           V0P=1.0D0+V0
           CALL GAMMA(V0P,GAP)
           A1=(0.5D0*X)**V0/GAP
        ENDIF
        K0=14
        IF (X.GE.35.0) K0=10
        IF (X.GE.50.0) K0=8
        IF (X.LE.18.0) THEN
           BI0=1.0D0
           R=1.0D0
           DO 15 K=1,30
              R=0.25D0*R*X2/(K*(K+V0))
              BI0=BI0+R
              IF (DABS(R/BI0).LT.1.0D-15) GO TO 20
15         CONTINUE
20         BI0=BI0*A1
        ELSE
           CA=DEXP(X)/DSQRT(2.0D0*PI*X)
           SUM=1.0D0
           R=1.0D0
           DO 25 K=1,K0
              R=-0.125D0*R*(VT-(2.0D0*K-1.0D0)**2.0)/(K*X)
25            SUM=SUM+R
           BI0=CA*SUM
        ENDIF
        M=MSTA1(X,200)
        IF (M.LT.N) THEN
           N=M
        ELSE
           M=MSTA2(X,N,15)
        ENDIF
        F2=0.0D0
        F1=1.0D-100
        DO 30 K=M,0,-1
           F=2.0D0*(V0+K+1.0D0)/X*F1+F2
           IF (K.LE.N) BI(K)=F
           F2=F1
30         F1=F
        CS=BI0/F
        DO 35 K=0,N
35         BI(K)=CS*BI(K)
        DI(0)=V0/X*BI(0)+BI(1)
        DO 40 K=1,N
40         DI(K)=-(K+V0)/X*BI(K)+BI(K-1)
        IF (X.LE.9.0D0) THEN
           IF (V0.EQ.0.0D0) THEN
              CT=-DLOG(0.5D0*X)-0.5772156649015329D0
              CS=0.0D0
              W0=0.0D0
              R=1.0D0
              DO 45 K=1,50
                 W0=W0+1.0D0/K
                 R=0.25D0*R/(K*K)*X2
                 CS=CS+R*(W0+CT)
                 WA=DABS(CS)
                 IF (DABS((WA-WW)/WA).LT.1.0D-15) GO TO 50
45               WW=WA
50            BK0=CT+CS
           ELSE
              V0N=1.0D0-V0
              CALL GAMMA(V0N,GAN)
              A2=1.0D0/(GAN*(0.5D0*X)**V0)
              A1=(0.5D0*X)**V0/GAP
              SUM=A2-A1
              R1=1.0D0
              R2=1.0D0
              DO 55 K=1,120
                 R1=0.25D0*R1*X2/(K*(K-V0))
                 R2=0.25D0*R2*X2/(K*(K+V0))
                 SUM=SUM+A2*R1-A1*R2
                 WA=DABS(SUM)
                 IF (DABS((WA-WW)/WA).LT.1.0D-15) GO TO 60
55               WW=WA
60            BK0=0.5D0*PI*SUM/DSIN(PIV)
           ENDIF
        ELSE
           CB=DEXP(-X)*DSQRT(0.5D0*PI/X)
           SUM=1.0D0
           R=1.0D0
           DO 65 K=1,K0
              R=0.125D0*R*(VT-(2.0*K-1.0)**2.0)/(K*X)
65            SUM=SUM+R
           BK0=CB*SUM
        ENDIF
        BK1=(1.0D0/X-BI(1)*BK0)/BI(0)
        BK(0)=BK0
        BK(1)=BK1
        DO 70 K=2,N
           BK2=2.0D0*(V0+K-1.0D0)/X*BK1+BK0
           BK(K)=BK2
           BK0=BK1
70         BK1=BK2
        DK(0)=V0/X*BK(0)-BK(1)
        DO 80 K=1,N
80         DK(K)=-(K+V0)/X*BK(K)-BK(K-1)
        VM=N+V0
        RETURN
        END


        SUBROUTINE GAMMA(X,GA)
C
C       ==================================================
C       Purpose: Compute gamma function a(x)
C       Input :  x  --- Argument of a(x)
C                       ( x is not equal to 0,-1,-2,uuu)
C       Output:  GA --- a(x)
C       ==================================================
C
        IMPLICIT DOUBLE PRECISION (A-H,O-Z)
        DIMENSION G(26)
        PI=3.141592653589793D0
        IF (X.EQ.INT(X)) THEN
           IF (X.GT.0.0D0) THEN
              GA=1.0D0
              M1=X-1
              DO 10 K=2,M1
10               GA=GA*K
           ELSE
              GA=1.0D+300
           ENDIF
        ELSE
           IF (DABS(X).GT.1.0D0) THEN
              Z=DABS(X)
              M=INT(Z)
              R=1.0D0
              DO 15 K=1,M
15               R=R*(Z-K)
              Z=Z-M
           ELSE
              Z=X
           ENDIF
           DATA G/1.0D0,0.5772156649015329D0,
     &          -0.6558780715202538D0, -0.420026350340952D-1,
     &          0.1665386113822915D0,-.421977345555443D-1,
     &          -.96219715278770D-2, .72189432466630D-2,
     &          -.11651675918591D-2, -.2152416741149D-3,
     &          .1280502823882D-3, -.201348547807D-4,
     &          -.12504934821D-5, .11330272320D-5,
     &          -.2056338417D-6, .61160950D-8,
     &          .50020075D-8, -.11812746D-8,
     &          .1043427D-9, .77823D-11,
     &          -.36968D-11, .51D-12,
     &          -.206D-13, -.54D-14, .14D-14, .1D-15/
           GR=G(26)
           DO 20 K=25,1,-1
20            GR=GR*Z+G(K)
           GA=1.0D0/(GR*Z)
           IF (DABS(X).GT.1.0D0) THEN
              GA=GA*R
              IF (X.LT.0.0D0) GA=-PI/(X*GA*DSIN(PI*X))
           ENDIF
        ENDIF
        RETURN
        END


        INTEGER FUNCTION MSTA1(X,MP)
C
C       ===================================================
C       Purpose: Determine the starting point for backward  
C                recurrence such that the magnitude of    
C                Jn(x) at that point is about 10^(-MP)
C       Input :  x     --- Argument of Jn(x)
C                MP    --- Value of magnitude
C       Output:  MSTA1 --- Starting point   
C       ===================================================
C
        IMPLICIT DOUBLE PRECISION (A-H,O-Z)
        A0=DABS(X)
        N0=INT(1.1*A0)+1
        F0=ENVJ(N0,A0)-MP
        N1=N0+5
        F1=ENVJ(N1,A0)-MP
        DO 10 IT=1,20             
           NN=N1-(N1-N0)/(1.0D0-F0/F1)                  
           F=ENVJ(NN,A0)-MP
           IF(ABS(NN-N1).LT.1) GO TO 20
           N0=N1
           F0=F1
           N1=NN
 10        F1=F
 20     MSTA1=NN
        RETURN
        END


        INTEGER FUNCTION MSTA2(X,N,MP)
C
C       ===================================================
C       Purpose: Determine the starting point for backward
C                recurrence such that all Jn(x) has MP
C                significant digits
C       Input :  x  --- Argument of Jn(x)
C                n  --- Order of Jn(x)
C                MP --- Significant digit
C       Output:  MSTA2 --- Starting point
C       ===================================================
C
        IMPLICIT DOUBLE PRECISION (A-H,O-Z)
        A0=DABS(X)
        HMP=0.5D0*MP
        EJN=ENVJ(N,A0)
        IF (EJN.LE.HMP) THEN
           OBJ=MP
           N0=INT(1.1*A0)
        ELSE
           OBJ=HMP+EJN
           N0=N
        ENDIF
        F0=ENVJ(N0,A0)-OBJ
        N1=N0+5
        F1=ENVJ(N1,A0)-OBJ
        DO 10 IT=1,20
           NN=N1-(N1-N0)/(1.0D0-F0/F1)
           F=ENVJ(NN,A0)-OBJ
           IF (ABS(NN-N1).LT.1) GO TO 20
           N0=N1
           F0=F1
           N1=NN
10         F1=F
20      MSTA2=NN+10
        RETURN
        END

        REAL*8 FUNCTION ENVJ(N,X)
        DOUBLE PRECISION X
        ENVJ=0.5D0*DLOG10(6.28D0*N)-N*DLOG10(1.36D0*X/N)
        RETURN
        END


      SUBROUTINE P3PGS ( A, B, N, FN, GN, VINT )

c*********************************************************************72
C
C  THIS SUBROUTINE USES THE PRODUCT TYPE THREE-POINT GAUSS-
C  LEGENDRE-SIMPSON RULE COMPOUNDED N TIMES TO APPROXIMATE
C  THE INTEGRAL FROM A TO B OF THE FUNCTION FN(X) * GN(X).
C  FN AND GN ARE FUNCTION SUBPROGRAMS WHICH MUST BE SUPPLIED
C  BY THE USER.  THE RESULT IS STORED IN VINT.
C
      DOUBLE PRECISION A, AG, AM(2,3), B, F(2), FN, G(3),
     &  GN, H, VINT, X(2), Y(2), DBLE

      DATA AM(1,1), AM(2,3) / 2 * 1.718245836551854D0 /,
     &  AM(1,2), AM(2,2) / 2 * 1.D0 /, AM(1,3), AM(2,1)
     &  / 2 * -.2182458365518542D0 /

      H = ( B - A ) / DBLE ( FLOAT ( N ) )
      X(1) = A + .1127016653792583D0 * H
      X(2) = A + .8872983346207417D0 * H
      Y(1) = A + H / 2.D0
      Y(2) = A + H
      VINT = 0.D0
      G(3) = GN ( A )
      DO 3 I = 1, N
        AG = FN ( Y(1) )
        G(1) = G(3)
        DO 1 J = 1, 2
          F(J) = FN ( X(J) )
          G(J+1) = GN ( Y(J) )
          X(J) = X(J) + H
1         Y(J) = Y(J) + H
        VINT = VINT + AG * 4.D0 * G(2)
        DO 3 J = 1, 2
          AG = 0.D0
          DO 2 K = 1, 3
2           AG = AG + AM(J,K) * G(K)
3         VINT = VINT + F(J) * AG
      VINT = H * VINT / 9.D0

      RETURN
      END



      function fn01(x)
      implicit none

      integer n

      double precision fn01
      double precision x

      fn01 = x

      return
      end

      function fn02(x)
      implicit none

      integer n

      double precision fn02
      double precision x

      fn02 = x**2.

      return
      end

      function gn01(x)
      implicit real*8(A-H,O-Z)

c      real*8 Psi_min,Psi_max,alpha_min,alpha_max,y,vint3
c      real*8 gamma

c      integer n
      external fn03
      external gn03      

c      double precision gn01
c      double precision x

      parameter (n=20)

      common gamma,y
    
      Psi_min = -asin(1./gamma) ! min vertical angle
      Psi_max = asin(1./gamma) ! max vertical angle
      alpha_min = 2.*gamma*Psi_min ! boosted min vertical angle
      alpha_max = 2.*gamma*Psi_max ! boosted max vertical angle
      alpha_cutoffm=-0.16
      alpha_cutoffp=0.16
      y=x
      call p3pgs(alpha_min,alpha_max,n,fn03,gn03,vint31)
c      call p3pgs(alpha_min,alpha_cutoffm,n,fn03,gn03,vint31)
c      call p3pgs(alpha_cutoffp,alpha_max,n,fn03,gn03,vint32)
      gn01 = vint31 !+vint32

      return
      end

      function gn02(x)
      implicit none

      real*8 Psi_min,Psi_max,alpha_min,alpha_max,y,vint4
      real*8 gamma

      integer n
      external fn04
      external gn04

      double precision gn02
      double precision x
      parameter (n=20)
      common gamma,y
    
      Psi_min = -asin(1./gamma) ! min vertical angle
      Psi_max = asin(1./gamma) ! max vertical angle
      alpha_min = 0. !16 !for the diff the int is from alpha_cutoff to alpha_max
      alpha_max = 2.*gamma*Psi_max ! boosted max vertical angle
      y=x
      call p3pgs(alpha_min,alpha_max,n,fn04,gn04,vint4)
      gn02 = vint4

      return
      end
 
      function fn03(x)
      implicit none


      integer n

      double precision fn03
      double precision x

    
      fn03 = (1+x**2.)**2.

      return
      end

      function fn04(x)
      implicit none


      integer n

      double precision fn04
      double precision x

    
      fn04 = x*(1+x**2.)**1.5

      return
      end

      function gn03(x)
      implicit real*8 (A-H,O-Z)

c      real*8 z,v,k23,k13,gamma,y,vm
c      dimension BI(0:*),DI(0:*),BK(0:*),DK(0:*)
  
c      integer n

      double precision gn03
      double precision x
      common gamma,y
      COMMON BI(0:250),DI(0:250),BK(0:250),DK(0:250)

      xk23=0.
      xk13=0.
      z=(y/2.)*(1+x**2)**1.5 ! z= (omega/2omega_c)*(1+alpha^2)^3/2
c      write(6,*)'gn03 gamma,y,x,z',gamma,y,x,z
      v=2./3.
      CALL IKV(V,z,VM,BI,DI,BK,DK)
      xk23=BK(0)

      v=1./3.
      CALL IKV(V,z,VM,BI,DI,BK,DK)
      xk13=BK(0)
c      write(6,*)'g03 k23 k13',xk23,xk13 
      gn03 = xk23**2. + x**2.*xk13**2/(1+x**2.)

      return
      end

      function gn04(x)
      implicit real*8(A-H,O-Z)

c      real*8 z,v,k23,k13,y,vm,gamma  
c      Dimension BI(0:*),DI(0:*),BK(0:*),DK(0:*)  

c      integer n

c      double precision gn04
c      double precision x
      Common gamma,y
      COMMON BI(0:250),DI(0:250),BK(0:250),DK(0:250)

      z=(y/2.)*(1+x**2)**1.5
      v=2./3.
      CALL IKV(V,z,VM,BI,DI,BK,DK)
      xk23=BK(0)

      v=1./3.
      CALL IKV(V,z,VM,BI,DI,BK,DK)
      xk13=BK(0)
 
      gn04 = xk23*xk13

      return
      end

\end{lstlisting}

\section{LANL Poisson SupeFish Geometry Description}
\lstset{language=Fortran, caption=LANL Poisson Input, label=bfield}
\begin{lstlisting}
Dipole Magnet Problem
DIPOLE 6 Simulation of Spin Light Chicane Magnet

; Copyright 1987, by the University of California.
; Unauthorized commercial use is prohibited.
; Author: Prajwal Mohanmurthy; Dec, 2011; (prajwal@jlab.org)

&reg kprob=0,       ! Poisson or Pandira problem
mode=0,             ! Some materials have variable permeability
nbsup=0,            ! added by gbf  !
nbslf=0,
rhogam=0.001        !try this gbf
xreg1=10.0,kreg1=82,! Physical and logical coordinates of x 
xreg2=25.0,kreg2=98,
xreg3=48.0,kreg3=150,

yreg1=14.0,lreg1=65,! Y line regions
yreg2=18.0,lreg2=70,
yreg3=21.0,lreg3=74,
lmax=80 &

&po x= -30.0000,y= 0.0000 &
&po x=60.000,y= 0.0000&
&po x=60.000,y=30.0000 &
&po x= -30.0000,y=30.0000 &
&po x= -30.0000,y= 0.0000 &

&reg mat=1,cur=-1040000.0 &
&po x=30.5000,y=6.000 &
&po x=40.0000,y=6.000 &
&po x=40.0000,y=12.000 &
&po x=30.5000,y=12.000 &
&po x=30.5000,y=6.000 &

&reg mat=1,cur=1040000.0 &
&po x=-10.0000,y=6.000 &
&po x=-0.5000,y=6.000 &
&po x=-0.5000,y=12.000 &
&po x=-10.000,y=12.000 &
&po x=-10.000,y=6.000 &

&reg mat=3,mtid=-2, mshape=0 &
&po x= 0.00,y=  5.62 &
&po x= 1.50,y=  2.62 &
&po x= 3.50,y=  1.27 &
&po x=26.50,y=  1.27 &
&po x=28.50,y=  2.65 &
&po x=30.00,y=  5.62 &
&po x=30.00,y= 12.50 &
&po x=40.50,y= 12.60 &
&po x=40.50,y=  0.00 &
&po x=60.00,y=  0.00 &
&po x=60.00,y= 30.00 &
&po x= 0.00,y= 30.00 &
&po x= 0.00,y=  5.62 &

&mt mtid=1
bgam=0.00000  0.0017513135
900     0.001747079
950     0.001741742
1000    0.001735498
1050    0.001728309
1100    0.00172014
1150    0.001710963
1200    0.001700753
1250    0.001689494
1300    0.001677174
1350    0.001663786
2800    0.001080694
2850    0.001068051
2900    0.001056142
2950    0.001044912
3000    0.001034309
3050    0.001024289
3100    0.001014809
3150    0.001005828
3200    0.000997312
3250    0.000989226
3300    0.000981539
3350    0.000974222
4000    0.000904952
4500    0.000856798
5000    0.000818493
5500    0.000788085
6000    0.000764202
6500    0.000745863
7000    0.000732376
7500    0.000723261
8000    0.000718209
8500    0.000717054
9000    0.000719758
9500    0.000726411
10000   0.000737231
10500   0.000752594
10578   0.0007562580
11319   0.0007951022
11940   0.0008375209
12451   0.0008834703
12912   0.0009293680
13313   0.0009764671
13654   0.0010253255
13935   0.0010764263
14216   0.0011254924
14447   0.0011767475
14618   0.0012313603
14789   0.0012846865
15020   0.0013315579
15131   0.0013879251
15252   0.0014423770
15432   0.0014912019
15594   0.0015389351
15705   0.0015918497
16180   0.0018542555
16840   0.0023752969
17150   0.0029154519
17360   0.0034566194
17620   0.0039729837
17830   0.0044863167
18200   0.0054945055
18950   0.0079176564
19500   0.0102564103
20200   0.0148588410
20650   0.0193798450
20950   0.0238663484
21600   0.0370370370
21900   0.0456621005
23000   0.0869565217
23386   0.1002810000
23850   0.1181630000
24408   0.1387420000
25079   0.1622460000
25885   0.1888580000
26854   0.2186950000
28019   0.2517840000 &

\end{lstlisting}

\section{Recursive SR Spectra Adding Code}
\lstset{language=Mathematica, caption=Mathematica7, label=box}
\begin{lstlisting}
!Recurssive SR - Spectra Adding Code
!Author: Prajwal Mohanmurthy, Dec 2011 (prajwal@jlab.org)
! Mathematica 7 File
up=30;
down=25;
asym=5;
sigup=up/E;
sigdown=down/E;
width=200; (*100+100*)
mid=width/2;
motion=10;
scale=0;
If[up-asym!= down,Print["Check Var : asym, down, up"]]
sr=Table[Table[0,{k,1,width+(2 motion)}],{kk,1,21}]; (*Set 'X' width =21*)
(*Print["-----------------"];*)
For[xx=1,xx<=21,xx++,
count=0;
count2=0;
x=xx-1;
For[y=0,If[x<=10,y<= x,y<=10-Abs[10-x] ],y++,
count++;
(*Print[x, ":x | y:",y];*)
For[k=motion+1+y,k<=motion+mid+y,k++,
sr[[xx]][[k]]+=Floor[(down*E^(-(((x-10)^2+(y)^2)/sigdown)))];
scale+=E^(-(((x-10)^2+(y)^2)/sigdown));
(*Print["Adding ",Floor[(down*E^(-(((x-10)^2+(y)^2)/sigdown)))]," to ",k];*)
];
For[k=motion+mid+1+y,k<=motion+width+y,k++,
sr[[xx]][[k]]+=Floor[(up*E^(-(((x-10)^2+(y)^2)/sigup )))];
scale+=E^(-(((x-10)^2+(y)^2)/sigdown));
(*Print["Adding ",Floor[(up*E^(-(((x-10)^2+(y)^2)/sigup )))]," to ",k];*)
];
];
(*Print["-----------------"];*)
For[y=0,If[x<= 10,y< x,y<10-Abs[10-x] ],y++,
count2++;
(*Print[x,  ":x | y:",-(y+1)];*)
For[k=motion+1+-(y+1),k<=motion+mid-(y+1),k++,
sr[[xx]][[k]]+=Floor[(down*E^(-(((x-10)^2+(y+1)^2)/sigdown)))];
scale+=E^(-(((x-10)^2+(y)^2)/sigdown));
(*Print["Adding ",Floor[(down*E^(-(((x-10)^2+(y+1)^2)/sigdown)))]," to ",k];*)
];
For[k=motion+mid+1-(y+1),k<=motion+width-(y+1),k++,
sr[[xx]][[k]]+=Floor[(up*E^(-(((x-10)^2+(y+1)^2)/sigup )))];
scale+=E^(-(((x-10)^2+(y)^2)/sigdown));
(*Print["Adding ",Floor[(up*E^(-(((x-10)^2+(y+1)^2)/sigup )))]," to ",k];*)
];
];
(*Print["-----------------"];*)
(*Print[count,"\t",count2,"\t",count+count2];*)
(*Print["-----------------"];*)
];
N[scale]
(*ListPlot3D[sr,PlotRange->Full]*)
400.023
nsize=520;
nwm=nsize-(2*motion);
Dimensions[sr];
sr2=Table[Table[0,{k,1,220}],{k,1,nsize}];
Dimensions[sr2]
Dimensions[sr2];
For[i2=1,i2<=21,i2++,
For[j2=1,j2<=220,j2++,
For[k2=i2,k2<nwm+i2,k2++,
sr2[[k2]][[j2]]+=sr[[i2]][[j2]];
];
];
];
Export["3dcontour_manypt.png",ListPlot3D[Transpose[sr2],ColorFunction->"BlueGreenYellow",AxesLabel->{"x","y","N"},PlotRange->Full]]
{520,220}
3dcontour_manypt.png
sr3=Table[Table[0,{k,1,220}],{k,1,nsize}];
Dimensions[sr3];
For[i2=11,i2<=110,i2++,
For[j2=11,j2<=nwm+motion,j2++,
sr3[[j2]][[i2]]=sr2[[nsize/4]][[50]];
];
];
For[i2=111,i2<=210,i2++,
For[j2=11,j2<=nwm+motion,j2++,
sr3[[j2]][[i2]]=Max[sr2];
];
];
Export["3dcontour_onept2.png",ListPlot3D[Transpose[sr3],AxesLabel->{"x","y","N"},ColorFunction->"BlueGreenYellow",PlotRange->Full]]
3dcontour_onept2.png
Clear[write2]
write2=OpenWrite["srXYdisc1.dat"]
WriteString[write2,"# x (micro m)" ,"|" ,"y (micro m)","|","N","\n"];
For[i=1,i<=220,i++,
For[j=1,j<=nsize,j++,
WriteString[write2,Floor[20 j] ,"\t" ,Floor[5  i],"\t",Floor[sr2[[j]][[i]]],"\n"];
];
]
Close[write2]
OutputStream[srXYdisc1.dat,36]
srXYdisc1.dat
Clear[write1]
write1=OpenWrite["srXYdiscone1.dat"]
WriteString[write1,"# x (micro m)" ,"|" ,"y (micro m)","|","N","\n"];
For[i=1,i<=220,i++,
For[j=1,j<=nsize,j++,
WriteString[write1,Floor[20 j] ,"\t" ,Floor[5  i],"\t",Floor[sr3[[j]][[i]]],"\n"];
];
]
Close[write1]
OutputStream[srXYdiscone1.dat,37]
srXYdiscone1.dat
\end{lstlisting}

\marginpar{For the complete program, refer to \url{mohanmurthy.com/a/SpinIC.gz.tar}}

\section{GEANT4 Geometry File}
\lstset{language=C++, caption=GEANT4 Toolkit - SpinIC executable, label=box}
\lstinputlisting{SpinICDetectorConstruction.cc}

\section{GEANT4 Stacking Action File}
\lstset{language=C++, caption=GEANT4 Toolkit - SpinIC executable, label=box}
\begin{lstlisting}
//////////////////////////////////////////////////////////////
// Author      : Prajwal Mohanmurthy, 09/2012               // 
//               (prajwal@mohanmurthy.com)                  //
//               Mississippi State University, MS 39762, USA//
// Description : Implements assymettry                      //
//               Counts SR + SL Photons vs. Energy          //        
//////////////////////////////////////////////////////////////   

#include "SpinICStackingAction.hh"
#include "SpinICAnalysis.hh"
#include "G4ProcessType.hh"
#include "G4VProcess.hh"
#include "G4Track.hh"
#include "Randomize.hh"
#include "G4Pow.hh"
#include <time.h>


SpinICStackingAction::SpinICStackingAction()
{
    G4cout << "********** Stacking Action Created *************" << G4endl;
}

SpinICStackingAction::~SpinICStackingAction()
{
    
}

G4ClassificationOfNewTrack SpinICStackingAction::ClassifyNewTrack(const G4Track * track)
{
    G4int PDG = track->GetParticleDefinition()->GetPDGEncoding();
    G4int parentID = track->GetParentID();
    G4int trackID = track->GetTrackID();
    G4int stepNum = track->GetCurrentStepNumber();

    G4int icnum = 3; //(1: Left IC, 2: right IC)
    G4int MagnetNumber = 0;
    G4ThreeVector Pos = track->GetPosition();

    G4double genergy = track->GetKineticEnergy()/MeV;
    G4double chrg = track->GetParticleDefinition()->GetPDGCharge();   
    G4double prodtm = track->GetGlobalTime();
    G4int prodstm = time(NULL);

    if (PDG == 22) {
        if (Pos.z()/m > -5.2 && Pos.z()/m < -4.9) {MagnetNumber = 1;}
        else if (Pos.z()/m > -6.85 && Pos.z()/m < -5.55) {MagnetNumber = 2;}
        else if (Pos.z()/m > -8.5 && Pos.z()/m < -8.2) {MagnetNumber = 3;}
        else {MagnetNumber = 5;}
    }

    // Get the creator process of the track that deposits energy
    const G4VProcess* edepCreatorProcess = track->GetCreatorProcess();
    G4ProcessType creatorProcessType = fNotDefined;
    G4int creatorProcessSubType = 0;
    if(edepCreatorProcess)
    {
        //edepCreatorProcess->DumpInfo();
        creatorProcessType = edepCreatorProcess->GetProcessType();
        creatorProcessSubType = edepCreatorProcess->GetProcessSubType();
    }
    G4int cpTypeCode = (G4int)creatorProcessType*10000 + creatorProcessSubType;

#if 1
    G4double phasymmetry = (4.543E-7) + (0.0000572046*genergy) + (-0.0000306751*std::pow(genergy,2)) + (0.0000138074*std::pow(genergy,3)) + (-3.37522E-6*std::pow(genergy,4)) + (3.29347E-7*std::pow(genergy,5));
    G4double asymmetry = 1. - phasymmetry;
    //G4double asymmetry = 1.; 
    
    //up-down asymmetry
    if (MagnetNumber==1 || MagnetNumber==3)
    {
        if (track->GetMomentumDirection().y() < 0. &&
        G4UniformRand() < asymmetry) {return fKill;}
    }
    else if (MagnetNumber==2)
    {
        if (track->GetMomentumDirection().y() > 0. &&
        G4UniformRand() < asymmetry) {return fKill;}
    }
    //Energy Spread asym
    if (G4UniformRand() > phasymmetry) {return fKill;}*/
#endif
    
    // Determine which IC based on xyz-position
      
    if (((Pos.z()/m >= 4.4) && (Pos.z()/m <= 6.6)) && ((Pos.y()/m >=-0.225) && (Pos.y()/m <= 0.225)))
    {
		 //icnum = 4;
         if ((Pos.x()/m >= -0.36) && (Pos.x()/m <= 0.0)){ icnum = 1; }
		 else if ((Pos.x()/m <= 0.36) && (Pos.x()/m >= 0.0)){ icnum = 2; }	
    }
       
    if(icnum==2 /*PDG==22 && genergy > 0.465 && parentID<=1000*/)
    {	
    // Get AnalysisManager to fill ntuples/histograms
    G4AnalysisManager* analysisManager = G4AnalysisManager::Instance();
        
    analysisManager->FillNtupleDColumn(0, genergy);
        
    analysisManager->FillNtupleDColumn(1, Pos.x()/m);
    analysisManager->FillNtupleDColumn(2, Pos.y()/m);
    analysisManager->FillNtupleDColumn(3, Pos.z()/m);
      
    analysisManager->FillNtupleDColumn(4, chrg);
    analysisManager->FillNtupleIColumn(5, PDG);
    analysisManager->FillNtupleIColumn(6, parentID);
    analysisManager->FillNtupleIColumn(7, cpTypeCode);
    //analysisManager->FillNtupleDColumn(8, prodtm);
    analysisManager->FillNtupleIColumn(8, MagnetNumber);
    analysisManager->FillNtupleIColumn(9, trackID);
        
    analysisManager->AddNtupleRow();
    }
    return fUrgent;
}
\end{lstlisting} 






\end{document}